\documentclass[12pt, draftclsnofoot, onecolumn]{IEEEtran}
\usepackage{times,amsmath,color,amssymb,graphicx}
\usepackage{mathrsfs}
\usepackage{amsfonts}
\usepackage{algorithm}
\usepackage{algorithmic}
\usepackage{epstopdf}
\usepackage{ntheorem}
\usepackage{subfigure}
\usepackage{booktabs}
\usepackage{cite}
\usepackage{float}
\theoremseparator{:}
\theorembodyfont{\rm}
\newtheorem{theorem}{\it \hskip 1em Theorem}

\long\def\symbolfootnote[#1]#2{\begingroup%
\def\thefootnote{\fnsymbol{footnote}}\footnote[#1]{#2}\endgroup}

\usepackage{multirow}

\def \beqi{\begin{IEEEeqnarray}{rcl}\IEEEyesnumber}
\def \eeqi{\end{IEEEeqnarray}}
\def \inum{\IEEEyessubnumber}

\newcommand\blfootnote[1]{%
\begingroup
\renewcommand\thefootnote{}\footnote{#1}%
\addtocounter{footnote}{-1}%
\endgroup
}

\begin{document}

\title{Joint Uplink-and-Downlink Optimization of 3D UAV Swarm Deployment for Wireless-Powered NB-IoT Networks}
\author{\IEEEauthorblockN{Han-Ting Ye, \emph{Student Member, IEEE}, Xin Kang, \emph{Senior Member, IEEE}, \\
Jingon Joung, \emph{Senior Member, IEEE}, Ying-Chang Liang, \emph{Fellow, IEEE}}
\vspace{-1cm}
}
\maketitle

%
%
%

\vspace{-0.6cm}
\begin{abstract}
This paper investigates a full-duplex orthogonal-frequency-division multiple access (OFDMA) based  multiple unmanned aerial vehicles (UAVs)-enabled wireless-powered Internet-of-Things (IoT) networks. In this paper, a swarm of UAVs is first deployed in three dimensions (3D) to simultaneously charge all devices, i.e., a downlink (DL) charging period, and then flies to new locations within this area to collect information from scheduled devices in several epochs via OFDMA due to potential limited number of channels available in Narrow Band IoT, i.e., an uplink (UL) communication period. To maximize the UL throughput of IoT devices, we jointly optimizes the UL-and-DL 3D deployment of the UAV swarm, including the device-UAV association, the scheduling order, and the UL-DL time allocation. In particular, the DL energy harvesting (EH) threshold of devices and the UL signal decoding threshold of UAVs are taken into consideration when studying the problem. Besides, both line-of-sight (LoS) and non-line-of-sight (NLoS) channel models are studied depending on the position of sensors and UAVs. The influence of the potential limited channels issue in NB-IoT is also considered by studying the IoT scheduling policy. Two scheduling policies, a near-first (NF) policy and a far-first (FF) policy, are studied. It is shown that the NF scheme outperforms FF scheme in terms of sum throughput maximization; whereas FF scheme outperforms NF scheme in terms of system fairness.
\end{abstract}
\vspace{-0.2cm}
\vspace{-1cm}
\blfootnote{Part of this work have been presented in \cite{9013224} at 2019 IEEE Global Communications Conference (GLOBECOM), Waikoloa, HI, USA, Dec 2019.}
\blfootnote{H.-T. Ye, X. Kang, and Y.-C. Liang are with Center for Intelligent Networking and Communications
(CINC), University of Electronic Science and Technology of
China (UESTC), Chengdu 611731, China (e-mail: yhtxyfs@gmail.com, kangxin@uestc.edu.cn, liangyc@ieee.org).}
\blfootnote{J. Joung is with the School of Electrical and Electronics Engineering, Chung-Ang
University, Seoul 06974, South Korea (e-mail: jgjoung@cau.ac.kr).}
\section{Introduction}
The Internet-of-Things (IoT) technology plays an important role in the upcoming era of big data, since it is necessary to perceive and capture enormous data through IoT embedded sensing devices \cite{ALAVI2018589}. According to \cite{hw}, the connection density of IoTs will reach to $1000\rm{k}~\rm{UEs/km^{2}}$, and a key factor affecting this scenario is the energy supply to the massive IoT devices. While the performance of processors and portable devices have been doubling every 18-24 months driven by Moore's law, battery technology in terms of capacity has only been growing with relatively low rate by $6\%$ per year. Even with power conscious designs and the latest in battery technology, many devices do not meet the lifetime cost and maintenance requirements for applications that require a large number of untethered devices, such as logistics and building automation. Today's devices performing two-way communication require scheduled-maintenance every three to 18 months to replace or recharge the power source (typically a battery) of the devices \cite{shearer2019powering}. One-way communicating devices that simply broadcast their status (one-way), such as automated utility meter readers, have a better battery life typically requiring the battery replacement within 10 years \cite{hw}. For both device types, scheduled power-source maintenance is costly and disruptive to monitor and/or control the entire system. Unscheduled maintenance trips are even more costly and disruptive. On a macro level, the relatively high cost associated with the internal battery also reduces the practical or economically viable number of devices that can be deployed.

The wireless power transfer (WPT) technology is undergoing rapid development because of its advantages, such as no contact, no wiring, reliable power supply, and the ease of maintenance. According to a recent report, wireless power transmission market is estimated to surge to 175 billions in 2027 \cite{WPT2}. The WPT technique mainly used in communication networks is the radio frequency (RF) energy transfer. It allows a longer effective charging distance (typically, within several tens of meters, up to several kilometers) and suitable for mobile applications \cite{7984754}. The RF microwave energy transfer also provides the advantages of immunity to the neighboring environment and the satisfactory of line-of-sight transfer requirement. As such, it is very suitable for powering a larger number of devices distributed in a wide area. In previous studies, low-power IoT networks mainly consist of ground sensors and hybrid access points (HAPs). HAPs provide stable energy to ground sensors and regularly collect data from them \cite{6800126,6678102,7115936}. However, the construction cost of HAPs is expensive. On the other hand, using unmanned aerial vehicle (UAV) as a moving HAP offers a more flexible and lower cost solution. For above reasons, UAV-enabled wireless-powered IoT networks have attracted great attention from researchers \cite{8489918,8632980,8761608,9080561}. In \cite{8489918}, throughput maximization problem was studied with a fixed-altitude UAV. In \cite{8632980}, the time resource and position of one UAV were jointly optimized to maximize the UL sum rate of all users. In \cite{8761608} and \cite{9080561}, a one dimensional-line model for UAV-enabled full-duplex IoT networks was studied.

However, the aforementioned studies have the following limitations: i) In reality, sensors can harvest energy only if the received signal strength is greater than a threshold, and decoders can successfully decode a signal if the signal to noise ratio (SNR) is larger than a threshold. However, these thresholds (namely, the threshold for downlink (DL) energy transfer and the threshold for uplink (UL) data transmission) were not considered in previous studies \cite{8489918,8632980,8761608,9080561}. ii) Line-of-sight (LoS) channels were mainly considered between the UAV and ground devices. However, in practice, the UAV-to-device communication link can be either LoS or non-line-of-sight (NLoS) according to the signal propagation environments \cite{6863654}. Recently, 3GPP released a technical report \cite{3GPP} describing the UAV channel model when synthesizing practical-measurement and ray-tracing simulations. It is important to note that there is an LoS probability in urban macro scenario when the UAV height is less than $100$ m. The probability of LoS/NLoS depends on the locations, heights, and the number of obstacles, as well as the elevation angle between the UAV and the associated ground devices. iii) The fixed altitude of the UAV was assumed in previous related works. In general, the altitude of the UAV can be flexibly adjusted so that the air-to-ground channel can be improved for better coverage. This is a key feature of UAV communications compared to traditional ground base station (BS) communications. iv) Time-division multiple access (TDMA) was generally assumed in previous UAV-enabled wireless powered IoT networks. However, OFDMA can be more efficient because it can improve the spectrum utilization.

Motivated by the issues and limitation of the existing studies above, in this paper, we investigate a full-duplex OFDMA (FD-OFDMA) based multiple UAVs-enabled wireless-powered IoT network, where a swarm of UAVs is deployed in three dimensional (3D) to simultaneously charge all devices and then fly to new locations to collect information from scheduled devices during several epochs via OFDMA. The main contribution of this study is listed as follows:
\begin{itemize}
\item We propose a new UAV-enabled 3D wireless-powered IoT model, which is different from existing models as follows: (i) The DL energy harvest (EH) threshold for devices and the UL SNR threshold for UAVs are taken into consideration when designing the system. (ii) We adopt a new channel model considering both LoS and NLoS channels \cite{6863654} of multiple altitude-adjustable UAVs. The altitude of each UAV can be dynamically adjusted to meet the UL and DL requirements considering the channel variation. (iii) A high efficient full-duplex OFDMA scheme is adopted in this system, under which we also study the scheduling policy to deal with the limited frequency resource issue in Narrow Band Internet of Things (NB-IoT).
\item Under the proposed model, we jointly optimize the UL-and-DL 3D deployment of the UAV swarm, including the device-UAV association, the IoT device scheduling order, and the UL-DL time allocation, to maximize the UL sum throughput. The proposed optimization problem is solved by investigating three sub-modules i) time allocation and scheduling optimization, ii) DL device association and UAV location optimization, and iii) UL device association and UAV location optimization.
\item We also propose two suboptimal scheduling strategies, referred to as the near-first (NF) scheme and far-first (FF) scheme, by exploiting the system characteristics. It is shown that the proposed suboptimal schemes can achieve a satisfactory performance. It is also observed that the NF scheme outperforms the FF scheme in terms of throughput maximization, but the FF scheme outperforms the NF scheme in terms of fairness.
\end{itemize}

The rest of this paper is organized as follows: Section \ref{System Model} describes the system model of the proposed 3D UAVs-enabled wireless-powered IoT networks, and the problem formulation of the sum throughput maximization problem. Section \ref{Sec-Solution} presents  the solution to the proposed optimization problem. Section \ref{Numerical Results} shows simulation results for the proposed algorithms and the comparison between the proposed schemes and conventional schemes. Section \ref{Sec-Conclusions} concludes this paper.

%
\vspace{-0.2cm}
\section{System Model and Problem Formulation} \label{System Model}
\begin{table}[t]
\setlength{\abovecaptionskip}{-0.1cm}   
\setlength{\belowcaptionskip}{-2cm}   
 \centering
 \caption{Notations}\label{tabel_PS}
\begin{tabular}{|c|c|c|}
\hline
\textbf{Symbols}&\textbf{Descriptions}\\
\hline
$K,N,M$&The numbers of devices, UAVs, and channels in network\\
$\mathcal{K},\mathcal{N}$&The set of all devices and UAVs defined as $\{1, 2, 3, \cdots, K\}\ni i$ and $\{1, 2, 3, \cdots, N\}\ni j$, respectively.\\
$\pmb{s}_{i},\pmb{u}_{j}\in \mathbb{R}^{3 \times 1}$&The location vector of each device $i$ and UAV $j$\\
$d_{i,j},\theta_{i,j},\bar{D}_{i,j}$&The distance, elevation angle, and average path loss between device $i$ and UAV $j$\\
$\tau_{0},\tau_{1}$&The time duration of DL and UL\\
$j^{D},j^{U}$&The indices of UAV $j$ at DL and UL\\
$\mathcal{L}_{j^{U}}$&The set of epochs for UAV $j^{U}$ in network defined as $\{1,2,\cdots,L_{j^{U}}\}\ni k$.\\
$\mathcal{Z}_{i}^{D},\mathcal{Z}_{i}^{U}$&The sets of UAVs that can successfully provide energy to device $i$ in DL and UL\\
$\mathcal{B}_{j^{D}}^{k}$&The set of devices charged by UAV $j^{D}$ in the epoch $k$\\
$\mathcal{C}_{j^{U}}^{k}$&The set of devices whose information is collected by UAV $j^{U}$ in the epoch $k$\\\
$s_{i,k}$&The indicator variable to denote whether the $i$th IoT device is scheduled for transmission in epoch $k$\\
$\mathcal{A}_{j^{U}}^{k}$&The set of devices served by UAV $j^{U}$ in the epoch $k$\\
$I_{i,j^{D}},a_{i,j^{U}},b_{i,j^{U}}$&The binary assignment variables in DL and UL, respectively.\\
$\textbf{I},\textbf{A},\textbf{B}\in \mathbb{R}^{K \times N}$&The assignment matrix corresponding to the variables $I_{i,j^{D}}$, $a_{i,j^{U}}$, $b_{i,j^{U}}$\\
\hline
\end{tabular}
\vspace{-0.2cm}
\end{table}

\begin{figure}[t]
\centerline{\includegraphics[width = 12cm]{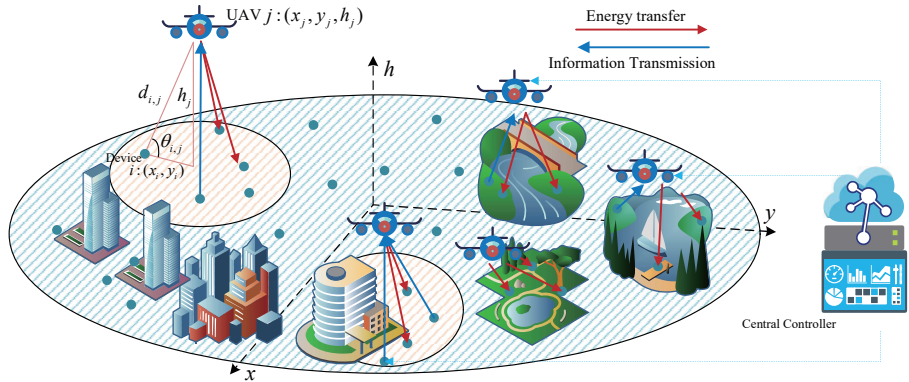}}
\vspace{-0.2cm}
\caption{System model.}
\vspace{-0.5cm}
\label{system_model_1}
\end{figure}

\begin{figure}[t]
\centering
\subfigure[DL: $\tau_{0}$  duration~~~~~~~~~~~~]{
\label{sa0}
\includegraphics[height=4.6cm]{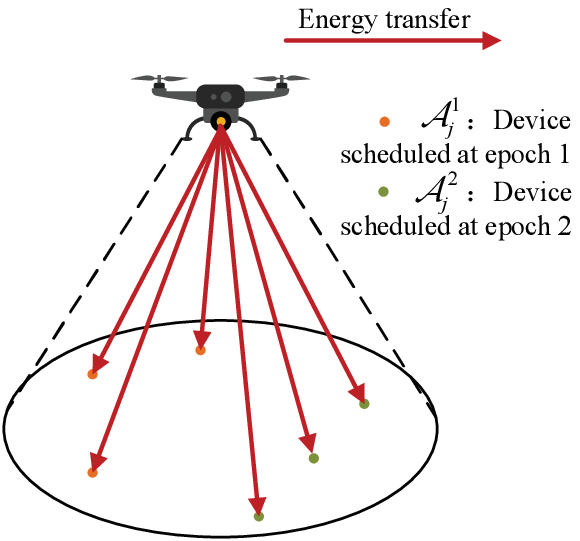}
}
\subfigure[UL: epoch 1 in $\tau_{1}$  duration.~~~~~~~~~~~~]{
\label{sa1}
\includegraphics[height=5cm]{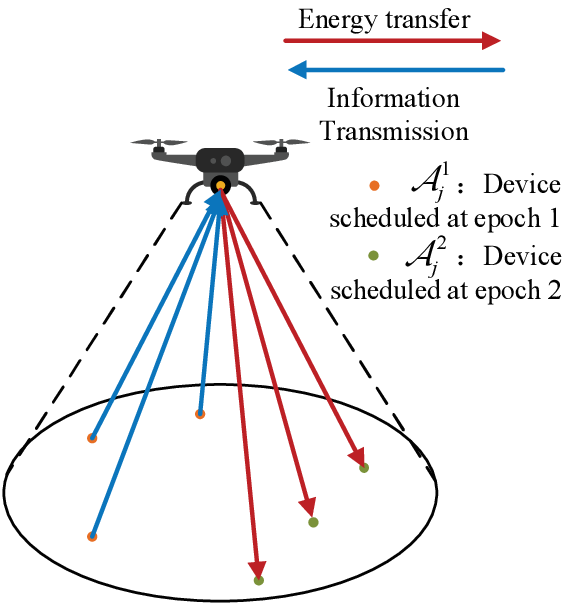}
}
\subfigure[UL: epoch 2 in $\tau_{1}$  duration.~~~~~~~~~~~~]{
\label{sa2}
\includegraphics[height=4.6cm]{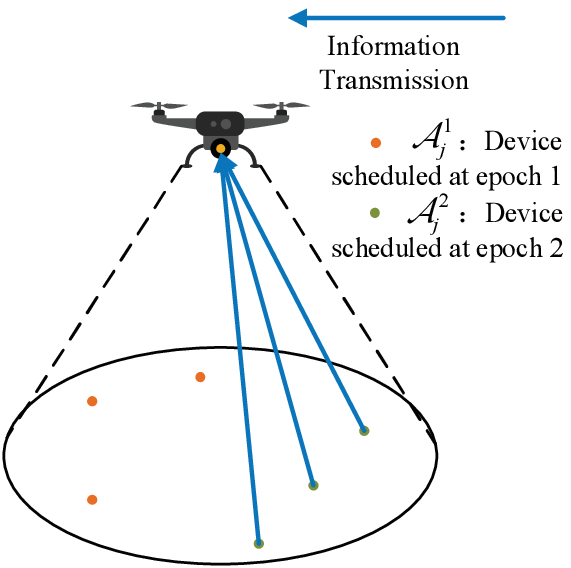}
}
\caption{Example illustration with six devices served by UAV $j$.} \label{sample_graph}
\vspace{-0.5cm}
\end{figure}

As shown in Fig. \ref{system_model_1}, a UAV-enabled wireless-powered IoT network is considered, in which $K$ single-antenna IoT devices are widely distributed and multiple two-antenna UAVs periodically charge and then collect data from ground IoT devices. In this work, we consider a centralized network, in which the locations of the devices and UAVs are known to a control center located at a central cloud server. Therefore, the server can calculate and obtain an accurate channel state information according to the positions of the UAVs and the positions of the devices and the previous channel state measurement information. The cloud server will then determine the downlink and uplink time allocation, the UAVs’ locations, the device-UAV association, and the scheduling order of each IoT device from the obtained channel state information. All UAVs operate in the full-duplex mode. For DL energy transfer, UAVs always continuously broadcast fixed energy signal $x_{ut}$ with a constant transmit power $P_{ut}$ to charge the devices that are located within a maximum wireless energy transfer range. Since $x_{ut}$ and the channel between the transmit and receive antennas known at UAV. The self-interference signal can be reconstructed at the receiving antenna and subtracted from the received signals. Therefore, the self-interference can be easily handled by using existing digital or analog cancellation techniques \cite{jain2011practical}. Once the devices harvest sufficient energy, they transmit information to the assigned UAV in the UL over orthogonal frequencies using the harvested energy. An simple example is given in Fig. \ref{sample_graph} to illustrate the procedure. A UAV first charges all devices in DL in Fig. \ref{sa0}. In the first epoch of UL in Fig. \ref{sa1}, the scheduled devices begin to transmit information to the UAV, whereas the devices scheduled at epoch 2 still harvest energy. During epoch 2 in Fig. \ref{sa2}, all the remaining devices transmit their information to the UAV. The UAVs can dynamically move to effectively serve the IoT devices during the DL energy transfer and UL data transmission. We establish a Cartesian coordinate system, where the center of the coverage area and the device location are donated by $(0,0,0)$ and $\pmb{s}_{i}=(x_{i},y_{i},0)^{T}\in \mathbb{R}^{3 \times 1},~\forall i \in \mathcal{K} \triangleq \{1, 2, 3, \cdots, K\}$, respectively. Let $\pmb{u}_{j}=(x_{j},y_{j},h_{j})^{T}\in \mathbb{R}^{3 \times 1}$  be the 3D coordinate of each UAV $j \in \mathcal{N} \triangleq \{1, 2, 3, \cdots, N\}$ with $h_{j}$ being the altitude of UAV $j$ as shown in Fig. \ref{system_model_1}. The location matrix for UAVs is $\textbf{L}_{N \times 3}=[\pmb{u}_{1}~ \pmb{u}_{2}~\cdots~\pmb{u}_{N}]$. For convenience, the most important variables used in this paper are defined in Table \ref{tabel_PS}.
\vspace{-0.5cm}
\subsection{Channel Model}
According to the 3GPP report \cite{3GPP}, the air-to-ground communication links can be either LoS or NLoS depending on the propagation environment and UAV altitude. For example, when a UAV is flying in a urban macro scenario at an altitude of less than 100 m, there will inevitably be LoS links and NLoS links, and the channel has a complete LoS condition when a UAV is higher than 100 m. In general, for a UAV-based communication system, complete information about exact locations, heights, and the number of obstacles may not be available \cite{8684899}. In this case, the randomness associated with LoS and NLoS links should be considered. The probability of having LoS communication links depends on locations, heights, and the number of obstacles, as well as the elevation angle between an UAV and its associated ground user. One suitable model for the LoS probability is given by \cite{6863654}
\begin{equation}\label{P_Los}
 Pr_{i,j}^{\rm{LoS}}=\left({1+\beta \exp(-\psi[\theta_{i,j}-\beta])}\right)^{-1},
\end{equation}
where $\psi$ and $\beta$ are constant values that depend on the carrier frequency and the type of environment (such as rural, urban, and dense urban) and $\theta_{i,j}$ is the elevation angle. Here, $\theta_{i,j}=\frac{180}{\pi}\arcsin(\frac{h_{j}}{d_{i,j}})$, where $d_{i,j}$ is the distance between device $i$ and UAV $j$ given as $d_{i,j}=\sqrt{(x_{j}-x_{i})^{2}+(y_{j}-y_{i})^{2}+h_{j}^{2}}$. From \eqref{P_Los}, it is evident that the LoS probability increases as either the elevation angle or the UAV altitude increases. The NLoS probability is then obtained as $Pr_{i,j}^{\rm{NLoS}}\!\!=\!\!1\!\!-\!\!Pr_{i,j}^{\rm{LoS}}$. Typically, given only the locations of the UAVs and devices, it is impossible to determine which path loss type (LoS/NLoS) is experienced between the node and UAV exactly. Therefore, we adopt the average channel gain, which has been widely used in 3D UAV-enabled wireless communication literature \cite{8053918,8038869}. Denoting the average path loss between device $i$ and UAV $j$ by $\bar{D}_{i,j}$, and the average channel gain between device $i$ and UAV $j$ can be expressed as
\begin{equation}\label{average_g}
\!\!g_{i,j}\!\!=\!\!\bar{D}_{i,j}^{-1}\!\!=\!\!(\kappa_{0}d_{i,j})^{-\alpha}\left(Pr_{i,j}^{\rm{LoS}}\mu^{\rm{LoS}}\!+\!Pr_{i,j}^{\rm{NLoS}}\mu^{\rm{NLoS}}\right)^{-1},
\end{equation}
where the factor $\kappa_{0}=4\pi f_{c}/c$ depends on carrier frequency $f_{c}$ and light speed $c$, and $\alpha$ is the path loss exponent. Herein, we set $\alpha\!\!=\!\!2$ for the LoS ground-to-air propagation \cite{6863654}. $\mu^{\rm{LoS}}$ and $\mu^{\rm{NLoS}}$ ($\mu^{\rm{NLoS}}\!\!>\!\!\mu^{\rm{LoS}}\!\!>\!\!1$) are the excessive path loss coefficients in LoS and NLoS cases, respectively. \eqref{average_g} can be explained as that in the air-to-ground link between the $i$th node and the $j$th UAV, $Pr_{i,j}^{\rm{LoS}}$ portion of the signals experience the LoS link type, and $Pr_{i,j}^{\rm{NLoS}}$ portion of the signals experience the NLoS link type. Note that, by using the average channel gain, there is no need to account for LoS and NLoS links separately, and hence, the throughput and SNR expressions become more tractable.

\vspace{-0.4cm}
\subsection{DL Energy Transfer}
Let $\pmb{\tau}=\{\tau_{0},\tau_{1}\}$ be two dynamic time slots for the proposed networks. All devices are charged by UAVs in the DL in the first time slot $\tau_{0}$, and all devices in the second time slot $\tau_{1}$ begin to transmit information to UAVs through the orthogonal channels. The indices of UAV $j$ at time $\tau_{0}$ and $\tau_{1}$ are denoted by $j^D$ and $j^U$, respectively. A position matrix of UAVs in the DL and UL are denoted by $\textbf{L}^{D}_{N \times 3}$ and $\textbf{L}^{U}_{N \times 3}$, respectively.

The energy harvested by device $i$ from UAV $j^{D}$ in $\tau_{0}$ is
\begin{equation}
E_{i,0}=\eta_{i}P_{ut}\tau_{0}\delta_{i}\sum_{j^{D}\in\mathcal{Z}_{i}^{D}}g_{i,j^{D}}(\resizebox{0.1\hsize}{!}{$d_{i,j^{D}},\theta_{i,j^{D}}$}),
\end{equation}
where $\delta_{i}$ indicates that only $\delta_{i}$ portion of multiple energy signals in free space successfully received by the device $i$ due to the constructive and destructive interferences; and $g_{i,j^{D}}(\resizebox{0.1\hsize}{!}{$d_{i,j^{D}},\theta_{i,j^{D}}$})$ denotes channel power gain which is a function of $d_{i,j^{D}}$ and $\theta_{i,j^{D}}$; and $\eta_i\in(0,1]$ denotes an EH efficiency factor for device $i$. Here, contrast of the conventional wireless powered networks, in which a high-power fixed HAP BS is prevented from over-estimating the amount of harvested power at the IoT devices \cite{7934322,8315145}, we consider a linear energy harvesting model in \cite{8489918,8632980,8761608,9080561} since the input power at the device is far from the nonlinear region when the RF signals reach the ground device through the air-to-ground channel.

Owing to the limited number of channels in the NB-IoT, denoted by $M$, we divide UL time $\tau_{1}$ into $L_{j^{U}}$ equal epochs, where $L_{j^{U}}=\lceil\frac{C_{j^{U}}}{M}\rceil$ and $C_{j^{U}}$ is the number of devices whose information is collected by UAV $j^{U}\in\mathcal{N}$. The set of epochs for UAV $j^{U}$ is defined as $\mathcal{L}_{j^{U}}=\{1,2,\cdots,L_{j^{U}}\}$. The length of each epoch is then $\frac{\tau_{1}}{L_{j^{U}}}$. Note that each UAV has two antennas to perform simultaneous DL energy transfer and the UL information transmission. This implies that devices scheduled later can harvest energy from the UAV in UL longer than the devices scheduled earlier. The energy harvested by device $i$ from UAVs in the UL during the $k$th epoch is
\begin{equation}
E_{i,k}=\eta_{i}P_{ut}\frac{(k-1)\tau_{1}}{L_{j^{U}}}\delta_{i}\sum_{j^{U}\in\mathcal{Z}_{i}^{U}}         g_{i,j^{U}}(\resizebox{0.1\hsize}{!}{$d_{i,j^{U}},\theta_{i,j^{U}}$}).
\end{equation}

To make our network practical, the following two constraints are considered in the system design.

\textbf{\textit{1) Energy harvesting constraints:}} The received power of the device must exceed a threshold, so that the ground devices can successfully harvest energy. To this end, the following two constraints should be satisfied, i.e.,
\beqi
&P_{ut}g_{i,j^{D}}(\resizebox{0.1\hsize}{!}{$d_{i,j^{D}},\theta_{i,j^{D}}$})\geq\rho, ~\forall j^{D} \in  \mathcal{Z}_{i}^{D},~\forall i \in  \mathcal{K},\inum\label{EH_constraint1}\\
&P_{ut}g_{i,j^{U}}(\resizebox{0.1\hsize}{!}{$d_{i,j^{U}},\theta_{i,j^{U}}$})\geq\rho, ~\forall j^{U} \in  \mathcal{Z}_{i}^{U},~\forall i \in  \mathcal{K},\inum\label{EH_constraint2}
\eeqi
where $\rho$ is the minimal RF input power required for ground devices \cite{6613706}, and $\mathcal{Z}_{i}^{D}$ and $\mathcal{Z}_{i}^{U}$ are the sets of UAVs that can successfully provide energy to device $i$ in DL and UL, respectively.

\textbf{\textit{2) DL service constraints:}} To ensure that each device is charged by at least one UAV, the following constraints are introduced:
\beqi
&\mathcal{Z}_{i}^{D}\neq\emptyset,~\mathcal{Z}_{i}^{D}\subset \mathcal{N},~\forall i \in \mathcal{K},\inum\label{DL_constraint1}\\
&\mathcal{Z}_{i}^{U}\neq\emptyset,~\mathcal{Z}_{i}^{U}\subset \mathcal{N},~\forall i \in \mathcal{K}.\inum\label{DL_constraint2}
\eeqi
\vspace{-1.5cm}
\subsection{UL Information Transmission}
First, we introduce an indicator to denote whether the $i$th IoT device is scheduled for transmission in epoch $k$ as follows:
\begin{equation}\label{s_ik}
s_{i,k}=\left\{
\begin{aligned}
&1,~~\mathrm{if}~i \in \mathcal{A}_{j^{U}}^{k},\\
&0,~~\mathrm{if}~i \notin \mathcal{A}_{j^{U}}^{k},
\end{aligned}
\right.
\end{equation}
where $\mathcal{A}_{j^{U}}^{k}$ represents the set of devices served by UAV $j^{U}$ in the epoch $k$. For UAV $j^{U}$, time slot $\tau_{1}$ is equally divided into $L_{j^{U}}$ epochs. Owing to the limited frequency resources in NB-IoT, only one frequency resource block is assumed to be allocated to each device. Thus, in each epoch, $M$ orthogonal resource blocks are allocated to $M$ different devices. Here, we consider that each UAV uses different frequency bands to prevent interference during the UL phase. We assume that the ground devices are equipped with super capacitor to store energy instead of battery, and thus whole harvested energy is used for transmission within its scheduled time slot.

In the UL communications, the following constraints need to be considered.

\textbf{\textit{1) SNR constraints:}} For the information receiver, i.e., the UAV, the signals can be recovered only if the received UL signals satisfy a certain SNR condition, which is
\begin{equation}\label{SNR_constraint}
\frac{\varepsilon_{i}[\tau_{0}L_{j^{U}}\!\!\sum_{j^{D}\in\mathcal{Z}_{i}^{D}}g_{i,j^{D}}(\resizebox{0.095\hsize}{!}{$d_{i,j^{D}},\theta_{i,j^{D}}$})\!\!+\!\!(k\!\!-\!\!1)\tau_{1}\!\!\sum_{j^{U}\in\mathcal{Z}_{i}^{U}}g_{i,j^{U}}(\resizebox{0.095\hsize}{!}{$d_{i,j^{U}},\theta_{i,j^{U}}$})]g_{i,j^{U}}(\resizebox{0.095\hsize}{!}{$d_{i,j^{U}},\theta_{i,j^{U}}$})}{\tau_{1}} \!\!\geq\!\! s_{i,k}\gamma,\! \forall i \! \in \! \mathcal{C}_{j^{U}}\!,
\end{equation}
where $\varepsilon_{i}$ is a constant defined as $\varepsilon_{i} \triangleq \eta_i\delta_{i}P_{ut}/N_{0}, ~\forall i \in \mathcal{K}$ and $\gamma$ is the SNR threshold. $N_{0}$ denotes the noise power at the UAV.

\textbf{\textit{2) UL service constraints:}} To ensure that each device delivers information to only one UAV, we introduce the following constraints:
\beqi
&\mathcal{C}_{m}\cap\mathcal{C}_{n}=\emptyset,~\forall m\neq n \in \mathcal{N},\inum\label{UL_constraint1}\\
&\bigcup_{j^{U}\in\mathcal{N}}\mathcal{C}_{j^{U}}=\mathcal{K}.\inum\label{UL_constraint2}
\eeqi

The transmission rate from device $i$ to UAV $j^{U}$ at the $k$th epoch is derived as follows:
\begin{align}
R_{i,j^{U}}^{k}&\!\!=\!\!\mathrm{ln} \!\!\left(1 + \frac{g_{i,j^{U}}(\resizebox{0.095\hsize}{!}{$d_{i,j^{U}},\theta_{i,j^{U}}$})(E_{i,0}+E_{i,k})}{\frac{\tau_{1}}{L_{j^{U}}} N_{0}} \right)\\
&\!\!=\!\!\mathrm{ln} \!\!\left(\!\!1 \!\!+\!\! \frac{\varepsilon_{i}[\tau_{0}L_{j^{U}}\sum_{j^{D}\in\mathcal{Z}_{i}^{D}}g_{i,j^{D}}(\resizebox{0.095\hsize}{!}{$d_{i,j^{D}},\theta_{i,j^{D}}$})\!\!+\!\!(k\!\!-\!\!1)\tau_{1}\sum_{j^{U}\in\mathcal{Z}_{i}^{U}}g_{i,j^{U}}(\resizebox{0.095\hsize}{!}{$d_{i,j^{U}},\theta_{i,j^{U}}$})]g_{i,j^{U}}(\resizebox{0.095\hsize}{!}{$d_{i,j^{U}},\theta_{i,j^{U}}$})}{\tau_{1}} \!\!\right)\!\!.\notag
\end{align}

Finally, the sum throughput of device $i$ served by UAV $j^{U}$ over $L_{j^{U}}$ epochs is given by
\begin{equation}
  \mathcal{T}_{i,j^{U}}=\sum_{k=1}^{L_{j^{U}}}s_{i,k}\frac{\tau_{1}}{L_{j^{U}}}R_{i,j^{U}}^{k}.\label{Throughput_i}
\end{equation}
\subsection{Problem Formulation}\label{Sec-PROBLEM}
In this paper, we are interested in finding the maximum throughput of the proposed network by considering a 3D deployment of UAV swarm, including jointly optimized 3D locations of UAVs, UAV-device associations, scheduling order, and time allocation. Note that, the UAV coordination is mainly reflected in the location distribution of multiple UAVs and the association between the UAV and the devices. Mathematically, the problem can be formulated as
\beqi
\!\!\!\!\!\!\!\!\underline{\textbf{OP:}}~~~~&\max_{\pmb{\tau},\pmb{s},\pmb{\mathcal{Z}}_{1\times K}^{D},\pmb{\mathcal{Z}}_{1\times K}^{U},\pmb{\mathcal{C}}_{1\times N},\textbf{L}^{D}_{N \times 3},\textbf{L}^{U}_{N \times 3}} &~\sum_{j^{U}=1}^{N}\sum_{i\in \mathcal{C}_{j^{U}}} \mathcal{T}_{i,j^{U}}\notag \label{1_determine}\\
&{\rm s.t.} &\!\!\!\!\!\!\!\!\!\!\!\!\!\!\!\!\!\!\!\!\!\!\!\!~\tau_{q} > 0,~\forall q \in  \{0,1\},  \inum\label{1_constraint_tau}\\
&&\!\!\!\!\!\!\!\!\!\!\!\!\!\!\!\!\!\!\!\!\!\!\!\!\tau_{0}+\tau_{1}\leq T_{\rm{hov}},\inum \label{1_constraint_sumtau}\\
&&\!\!\!\!\!\!\!\!\!\!\!\!\!\!\!\!\!\!\!\!\!\!\!\!s_{i,k}\in\{0,1\}, ~\forall i \in  \mathcal{C}_{j^{U}},~\forall k \in  \mathcal{L}_{j^{U}},~\forall j^{U} \in  \mathcal{N}, \inum \label{1_constraint_sik}\\
&&\!\!\!\!\!\!\!\!\!\!\!\!\!\!\!\!\!\!\!\!\!\!\!\!\sum_{i\in\mathcal{C}_{j^{U}}}s_{i,k}\leq M, \sum_{k \in \mathcal{L}_{j^{U}}}s_{i,k}=1,  ~\forall i \in  \mathcal{C}_{j^{U}}, ~\forall k \in \mathcal{L}_{j^{U}},~\forall j^{U} \in \mathcal{N}, \inum \label{1_constraint_sums}\\
&&\!\!\!\!\!\!\!\!\!\!\!\!\!\!\!\!\!\!\!\!\!\!\!\!\eqref{EH_constraint1},\eqref{EH_constraint2},\eqref{DL_constraint1},\eqref{DL_constraint2},\eqref{SNR_constraint},\eqref{UL_constraint1} ,~\mbox{and}~ \eqref{UL_constraint2},\notag
\eeqi
where $\pmb{\mathcal{Z}}^{D}_{1\times K}=[\mathcal{Z}_{1}^{D}~\mathcal{Z}_{2}^{D}~\cdots~\mathcal{Z}_{K}^{D}]$ and  $\pmb{\mathcal{Z}}^{U}_{1\times K}=[\mathcal{Z}_{1}^{U}~\mathcal{Z}_{2}^{U}~\cdots~\mathcal{Z}_{K}^{U}]$ are vectors for EH service sets in DL and UL, respectively. $\pmb{\mathcal{C}}_{1\times N}=[\mathcal{C}_{1}~\mathcal{C}_{2}~\cdots~\mathcal{C}_{N}]$ is a vector for the UL service set. The constraints of \eqref{1_constraint_tau} and \eqref{1_constraint_sumtau} represent non-negativity of time, and the total time constraint, respectively. $T_\mathrm{hov}$ is the total hovering time for time slots in DL and UL. Energy consumption is an important feature of UAVs \cite{9080561,9042882,8941314}, however, a short duration of movement of UAVs between DL and UL is not considered in this study. Note that the total time constraint \eqref{1_constraint_sumtau} has a similar effect with the total energy constraint. For the energy consumption of the intermediate movement process, it can be modeled separately so that the energy consumed by the movement is minimized, which is put as our future work. For convenience, we use a normalized unit block time, i.e., $T_{\rm{hov}} = 1$. The constraints \eqref{1_constraint_sik} and \eqref{1_constraint_sums} indicate that each device can be scheduled at most one epoch and at most $M$ devices perform UL information transmission at each epoch due to NB-IoT's limited number of channels.

In \textbf{OP}, constraints \eqref{EH_constraint1}, \eqref{EH_constraint2}, and \eqref{SNR_constraint} are nonlinear and nonconvex due to the non-convexity \eqref{P_Los}. Moreover, the DL constraints \eqref{EH_constraint1}, \eqref{EH_constraint2}, \eqref{DL_constraint1}, and \eqref{DL_constraint2} are coupled with UL constraints \eqref{SNR_constraint},\eqref{UL_constraint1}, and \eqref{UL_constraint2}. Besides, DL locations of UAVs and device-UAV association have an impact on UL locations of UAVs and device-UAV association due to the causality of energy harvest communication. For all these reasons, the proposed \textbf{OP} is not a convex optimization problem, and thus can not be readily solved by existing convex optimization algorithms.
\section{Solution to the Proposed Optimization Problem}\label{Sec-Solution}
\begin{figure}[t]
\centerline{\includegraphics[width = 14cm]{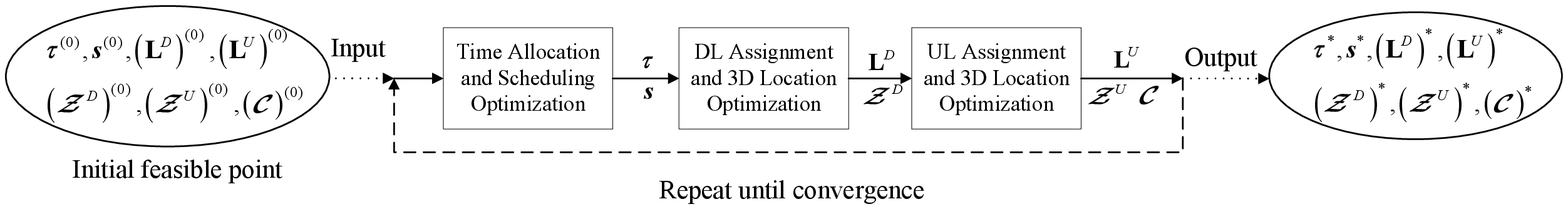}}
\vspace{-0.2cm}
\caption{Optimization procedures for the proposed solution.}
\vspace{-0.5cm}
\label{Block_digram}
\end{figure}

For the above reasons, in this section, we propose a novel framework to solve \textbf{OP}. Fig. \ref{Block_digram} shows a block diagram that summarizes the main steps for solving \textbf{OP}. First, given the locations of UAVs and associations between the UAVs and devices in DL and UL, we derive the optimal time allocation and device scheduling. Next, with the optimal time allocation and fixed scheduling obtained from the previous step, we transform \textbf{OP} to a mixed-integer sum-of-ratios problem, and decompose it into the linear 0-1 fractional programming and nonlinear fractional programming problems to design the locations of UAVs and device-UAV associations in DL. Finally, given the other variables obtained from the previous two optimization steps, the locations of UAVs and associations between the UAVs and devices in UL can be obtained from a sum-of-ratios problem, which can be solved by using a Dinkelbach-based algorithm. The above procedure is performed iteratively until all solutions are converged. Next, we discuss each step to obtain the solutions in detail.
\subsection{Time allocation and scheduling optimization}\label{Sec-TIME}
For the given locations of UAVs and user associations, namely $\textbf{L}^{D}$, $\textbf{L}^{U}$, $\pmb{\mathcal{Z}}^{D}$, $\pmb{\mathcal{Z}}^{U}$ and $\pmb{\mathcal{C}}$, the original problem \textbf{OP} can be converted into a convex optimization problem as follows:
\beqi\label{FW_subP1}\underline{\textbf{P1:}}~~~~~~~~~
&\max_{\pmb{\tau},\pmb{s}} &~\sum_{j^{U}=1}^{N}\sum_{i=1}^{C_{j^{U}}}\sum_{k=1}^{L_{j^{U}}}s_{i,k}\frac{\tau_{1}}{L_{j^{U}}}\mathrm{ln} \left(1 + \frac{\Theta_{i}^{0}L_{j^{U}}\tau_{0}+\Theta_{i}^{1}(k-1)\tau_{1}}{\tau_{1}} \right)\inum \label{11_determine}\\
&{\rm s.t.} &~\eqref{1_constraint_tau},\eqref{1_constraint_sumtau}, \eqref{1_constraint_sik},~\mbox{and}~\eqref{1_constraint_sums}\notag\\
&&\frac{\tau_{1}s_{i,k}\gamma}{\Theta_{i}^{0}L_{j^{U}}\tau_{0}+\Theta_{i}^{1}(k-1)\tau_{1}}\leq1, ~\forall i \in  \mathcal{C}_{j^{U}},~\forall j^{U} \in  \mathcal{N}, \inum\label{11_constraint_SNR}
\eeqi
where $C_{j^{U}}=|\mathcal{C}_{j^{U}}|$ is the number of devices assigned to UAV $j^{U}$, and \eqref{11_constraint_SNR} is the simplification of SNR constraint \eqref{SNR_constraint} when the locations of UAVs and user associations in DL and UL are given. $\Theta_{i}^{0}\!\!=\!\!\varepsilon_{i}g_{i,j^{U}}(\!\resizebox{0.095\hsize}{!}{$d_{i,j^{U}},\theta_{i,j^{U}}$}\!)\sum_{j^{D}\in\mathcal{Z}_{i}^{D}}g_{i,j^{D}}(\!\resizebox{0.095\hsize}{!}{$d_{i,j^{D}},\theta_{i,j^{D}}$}\!)$ and $\Theta_{i}^{1}\!\!=\!\!\varepsilon_{i}g_{i,j^{U}}(\!\resizebox{0.095\hsize}{!}{$d_{i,j^{U}},\theta_{i,j^{U}}$}\!)\sum_{j^{U}\in\mathcal{Z}_{i}^{U}}g_{i,j^{U}}(\!\resizebox{0.095\hsize}{!}{$d_{i,j^{U}},\theta_{i,j^{U}}$}\!)$.

For \textbf{P1}, we can derive the following theorem.

\begin{theorem} \label{theorm-1-time}
Vector $\pmb{\tau}^{*}=[\tau_0^*,\tau_1^*]$ is the optimal time allocations of \textbf{P1} for all UAVs, when it satisfy the following:
\begin{equation}\label{opt_tau}
\pmb{\tau}^{*}=\left\{
\begin{aligned}
&\mathrm{Root}\left(\sum_{j^{U}=1}^{N}\sum_{i=1}^{C_{j^{U}}}\sum_{k=1}^{L_{j^{U}}}s_{i,k}^{*} R_{i,j^{U}}^{k} = \sum_{j=1}^{N}\sum_{i=1}^{C_{j^{U}}}\sum_{k=1}^{L_{j^{U}}} \frac{s_{i,k}^{*} \Theta_{i}^{0}L_{j^{U}}}{\Theta_{i}^{0}L_{j^{U}}\tau_{0}^{*}+[1+\Theta_{i}^{1}(k-1)]\tau_{1}^{*} }\right),~~\varpi<1,\\
&\left[\frac{\gamma-\Theta_{m}^{1}(n-1)}{\gamma+\Theta_{m}^{0}L_{j^{U}}-\Theta_{m}^{1}(n-1)},
  \frac{\Theta_{m}^{0}L_{j^{U}}}{\gamma+\Theta_{m}^{0}L_{j^{U}}-\Theta_{m}^{1}(n-1)}\right],~~\varpi\geq1,
\end{aligned}
\right.
\end{equation}
where $\varpi=\frac{\tau_{1}^{*}s_{m,n}^{*}\gamma}{\Theta_{m}^{0}L_{j^{U}}\tau_{0}^{*}+\Theta_{m}^{1}(n-1)\tau_{1}^{*}}$ and $\{m,n\}=\arg\max_{\{i \in  \mathcal{K}, k\in \mathcal{L}_{j^{U}}\}}\frac{\tau_{1}^{*}s_{i,k}^{*}\gamma}{\Theta_{i}^{0}L_{j^{U}}\tau_{0}^{*}+\Theta_{i}^{1}(k-1)\tau_{1}^{*}}$, and the operator $\mathrm{Root}\left(\cdot\right)$ finds the roots of an equation, and the set of devices served by UAV $j^{U}$ at epoch $k$ is obtained by as follows:
\begin{equation}\label{theorem-A_juk}
\mathcal{A}_{j^{U}}^{k} \!\!=\!\!\arg\!\!\!\!\!\!\!\!\!\!\max_{\substack{\mathcal{C}_{j^{U}}^{(S)}\subset\left(\mathcal{C}_{j^{U}}/\sum_{n=k+1}^{L_{j^{U}}}{A}_{j^{U}}^{n}\right), \\
|\mathcal{C}_{j^{U}}^{(S)}|\leq M}}\!\!\sum_{i\in \mathcal{C}_{j^{U}}^{(S)}}\!\!\frac{\tau_{1}^{*}}{L_{j^{U}}}\mathrm{ln}\! \left(\!1 \!+\! \frac{\Theta_{i}^{0}L_{j^{U}}\tau_{0}^{*}\!+\!\Theta_{i}^{1}(k\!-\!1)\tau_{1}^{*}}{\tau_{1}^{*}} \!\right)\!-\!\frac{w_{i,k}^{*}\tau_{1}^{*}\gamma}{\Theta_{i}^{0}L_{j^{U}}\tau_{0}^{*}+\Theta_{i}^{1}(k\!-\!1)\tau_{1}^{*}}.
\end{equation}
Here, $w_{i,k}^{*}$ is obtained as follows:
\begin{equation}\label{theorem-w_ik}
w_{i,k}^{*}=\left\{
\begin{aligned}
&1,~\frac{\tau_{1}^{*}\gamma}{\Theta_{i}^{0}L_{j^{U}}\tau_{0}^{*}+\Theta_{i}^{1}(k\!-\!1)\tau_{1}^{*}}\geq1,\\
&0,~\frac{\tau_{1}^{*}\gamma}{\Theta_{i}^{0}L_{j^{U}}\tau_{0}^{*}+\Theta_{i}^{1}(k\!-\!1)\tau_{1}^{*}}<1.
\end{aligned}
\right.
\end{equation}
\end{theorem}
\begin{IEEEproof}
Please refer to Appendix \ref{proof-1-time}.
\end{IEEEproof}

An interesting point of Theorem \ref{theorm-1-time} is that we can find that the optimal solution of \textbf{P1} that makes the sum rate of UL communication in the DC phase an optimum value. In addition, by comparing \eqref{opt_tau} and the objective function \eqref{11_determine} in \textbf{P1}, we can convert the logarithmic form in \eqref{11_determine} into a fractional expression.

\subsection{DL assignment and 3D location optimization}\label{Sec-DL}
For the given optimal time allocation, UL scheduling, UL device association, and UAV locations, we optimize the DL assignment and 3D locations of UAVs. Based on Theorem \ref{theorm-1-time}, the optimization problem \textbf{P1} under condition that $\varpi<1$ can be reformulated as follows:
\begin{align}
  \max_{\textbf{L}^{D},\pmb{\mathcal{Z}}^{D}} \sum_{j^{U}=1}^{N}\sum_{i=1}^{C_{j^{U}}}\sum_{k=1}^{L_{j^{U}}} \frac{\tau_{1}^{*}}{L_{j^{U}}}R_{i,j^{U}}^{k} \Longleftrightarrow&\!\max_{\textbf{L}^{D},\pmb{\mathcal{Z}}^{D}} \!\sum_{j^{U}=1}^{N}\sum_{i=1}^{C_{j^{U}}}\sum_{k=1}^{L_{j^{U}}} \frac{s_{i,k}^{*}\Theta_{i}^{0}L_{j^{U}}}{\Theta_{i}^{0}L_{j^{U}}\tau_{0}^{*}+[1+\Theta_{i}^{1}(k-1)]\tau_{1}^{*} }, \\
  \Longleftrightarrow&\!\max_{\textbf{L}^{D},\pmb{\mathcal{Z}}^{D}} \!\sum_{j^{U}=1}^{N}\sum_{i=1}^{C_{j^{U}}}\sum_{k=1}^{L_{j^{U}}}\frac{s_{i,k}^{*}}{\Phi_{i}[1\!+\!\Theta_{i}^{1}(k\!-\!1)]/\sum_{j^{D}\in\mathcal{Z}_{i}^{D}}g_{i,j^{D}}(\resizebox{0.08\hsize}{!}{$d_{i,j^{D}},\theta_{i,j^{D}}$}) \!+\!  \tau_{0}^*},\notag\\
  \Longleftrightarrow&\!\min_{\textbf{L}^{D},\pmb{\mathcal{Z}}^{D}} \!\sum_{j^{U}=1}^{N}\sum_{k=1}^{L_{j^{U}}}\sum_{i=1}^{A_{j^{U}}^{k}}\frac{\Phi_{i}[1\!+\!\Theta_{i}^{1}(k\!-\!1)]\!-\!\tau_{1}^*\sum_{j^{D}\in\mathcal{Z}_{i}^{D}}g_{i,j^{D}}(\resizebox{0.095\hsize}{!}{$d_{i,j^{D}},\theta_{i,j^{D}}$})}{\Phi_{i}[1\!+\!\Theta_{i}^{1}(k\!-\!1)] \!+\!  \tau_{0}^*\sum_{j^{D}\in\mathcal{Z}_{i}^{D}}g_{i,j^{D}}(\resizebox{0.095\hsize}{!}{$d_{i,j^{D}},\theta_{i,j^{D}}$})},\notag
\end{align}
where $\Phi_{i}=\frac{\tau_{1}^{*}}{\varepsilon_{i}L_{j^{U}}g_{i,j^{U}}(\resizebox{0.095\hsize}{!}{$d_{i,j^{U}},\theta_{i,j^{U}}$})}$.

Subsequently, the optimization problem \textbf{P1} under condition that $\varpi\geq1$ can be reformulated as follows:
\begin{align}
  \!\!\!\!\max_{\textbf{L}^{D},\pmb{\mathcal{Z}}^{D}} \!\sum_{j^{U}=1}^{N}\sum_{i=1}^{C_{j^{U}}}\sum_{k=1}^{L_{j^{U}}} \frac{\tau_{1}^{*}}{L_{j^{U}}}R_{i,j^{U}}^{k} \!\Longleftrightarrow&\!\max_{\textbf{L}^{D},\pmb{\mathcal{Z}}^{D}} \sum_{j^{U}=1}^{N}\sum_{i=1}^{C_{j^{U}}}\sum_{k=1}^{L_{j^{U}}} s_{i,k}^{*}\ln\!\left(\!1\!+\!\frac{\Theta_{i}^{0}\gamma\!-\!\Theta_{i}^{0}\Theta_{m}^{1}(n\!-\!1)\!+\!\Theta_{i}^{1}(k\!-\!1)\Theta_{m}^{0}}{\Theta_{m}^{0}}\!\right), \notag\\
  \overset{(a)}\Longleftrightarrow&\!\min_{\textbf{L}^{D},\pmb{\mathcal{Z}}^{D}} \sum_{j^{U}=1}^{N}\sum_{i=1}^{C_{j^{U}}}\sum_{k=1}^{L_{j^{U}}} \frac{s_{i,k}^{*}\Theta_{m}^{0}}{\Theta_{i}^{0}[\gamma\!-\!\Theta_{m}^{1}(n\!-\!1)]\!+\![1\!+\!\Theta_{i}^{1}(k\!-\!1)]\Theta_{m}^{0}},\\
  \Longleftrightarrow&\!\min_{\textbf{L}^{D},\pmb{\mathcal{Z}}^{D}} \sum_{j^{U}=1}^{N}\sum_{k=1}^{L_{j^{U}}}\sum_{i=1}^{A_{j^{U}}^{k}} \frac{1}{\Gamma_{i}\sum_{j^{D}\in\mathcal{Z}_{i}^{D}}g_{i,j^{D}}(\resizebox{0.095\hsize}{!}{$d_{i,j^{D}},\theta_{i,j^{D}}$})\!+\!\Theta_{i}^{1}(k\!-\!1)\!+\!1},\notag
\end{align}
where `$(a)$' come from the fact that $\mathrm{ln}x \geq 1-1/x$ for $x>0$ and $\Gamma_{i}=\frac{\varepsilon_{i}g_{i,j^{U}}(\resizebox{.01\hsize}{!}{\textit{d}}_{i,j^{U}},\theta_{i,j^{U}})[\gamma-\Theta_{m}^{1}(n-1)]}{\Theta_{m}^{0}}$.

In this section, we thus mainly discuss the optimization of the following function, i.e.,
\begin{equation}\label{sumR}
F_{2}=\left\{
\begin{aligned}
&\sum_{j^{U}=1}^{N}\sum_{k=1}^{L_{j^{U}}}\sum_{i=1}^{A_{j^{U}}^{k}}\frac{\Phi_{i}[1+\Theta_{i}^{1}(k-1)]-\tau_{1}^*\sum_{j^{D}\in\mathcal{Z}_{i}^{D}}g_{i,j^{D}}(\resizebox{0.095\hsize}{!}{$d_{i,j^{D}},\theta_{i,j^{D}}$})}{\Phi_{i}[1+\Theta_{i}^{1}(k-1)] +  \tau_{0}^*\sum_{j^{D}\in\mathcal{Z}_{i}^{D}}g_{i,j^{D}}(\resizebox{0.095\hsize}{!}{$d_{i,j^{D}},\theta_{i,j^{D}}$})},~~\varpi<1,\\
&\sum_{j^{U}=1}^{N}\sum_{k=1}^{L_{j^{U}}}\sum_{i=1}^{A_{j^{U}}^{k}} \frac{1}{\Gamma_{i}\sum_{j^{D}\in\mathcal{Z}_{i}^{D}}g_{i,j^{D}}(\resizebox{0.095\hsize}{!}{$d_{i,j^{D}},\theta_{i,j^{D}}$})+\Theta_{i}^{1}(k-1)+1},~~\varpi\geq1.\
\end{aligned}
\right.
\end{equation}

Given the optimal time allocation, UL scheduling, UL 3D locations of UAVs and UL association between UAVs and devices, we optimize the DL assignment and 3D locations of UAVs.
\beqi\!\!\!\!\!\!\!\!\underline{\textbf{P2:}}~~~~
&\min_{\textbf{L}^{D},\pmb{\mathcal{Z}}^{D}}&~F_{2},\inum \label{1b_determine}\\
&{\rm s.t.} &~\eqref{EH_constraint1},~\mbox{and}~\eqref{DL_constraint1}.\notag
\eeqi

Directly solving \textbf{P2} is challenging, because the UAVs' locations and device association are coupled. In particular, to obtain the device association, the locations of the UAVs must be known. Moreover, the UAVs' locations cannot be optimized without knowing the device association. Therefore, we decompose \textbf{P2} into two subproblems. In the first subproblem, given the fixed locations of UAVs, the DL devices associations are optimized. In the second subproblem, given the optimal device association from the first subproblem, the suboptimal DL 3D locations of UAVs are designed to maximize the sum throughput of the devices.

\subsubsection{DL Assignment Step}
First, we introduce binary variable $I_{i,j^{D}}$, where $I_{i,j^{D}}=1$ if device $i$ can harvest energy from radio signals from UAV $j^{D}$, $I_{i,j^{D}}=0$ otherwise. And the corresponding allocation matrix is $\textbf{I}_{K \times N}$. The set of UAVs which covers device $i$ is then defined as follows:
\begin{equation}
  \mathcal{Z}_{i}^{D}\triangleq\{n|I_{i,n}=1, ~\forall n \in  \mathcal{N}\},~\forall i \in  \mathcal{K}.
\end{equation}
Using $I_{i,j^{D}}$, $F_{2}$ can be rewritten as follows:
\begin{equation}\label{sumR}
F_{2a}=\left\{
\begin{aligned}
&\frac{\Phi_{i}[1+\Theta_{i}^{1}(k-1)]-\tau_{1}^*\sum_{j^{D}=1}^{N}I_{i,j^{D}}/\bar{D}_{i,j^{D}}(\resizebox{0.095\hsize}{!}{$d_{i,j^{D}},\theta_{i,j^{D}}$})}{\Phi_{i}[1+\Theta_{i}^{1}(k-1)] +  \tau_{0}^*\sum_{j^{D}=1}^{N}I_{i,j^{D}}/\bar{D}_{i,j^{D}}(\resizebox{0.095\hsize}{!}{$d_{i,j^{D}},\theta_{i,j^{D}}$})},~~\varpi<1\\
&\frac{1}{\Gamma_{i}\sum_{j^{D}=1}^{N}I_{i,j^{D}}/\bar{D}_{i,j^{D}}(\resizebox{0.095\hsize}{!}{$d_{i,j^{D}},\theta_{i,j^{D}}$})+\Theta_{i}^{1}(k-1)+1},~~\varpi\geq1\
\end{aligned}
\right.,~\forall i \in \mathcal{A}_{j^{U}}^{k},~\forall k \in \mathcal{L}_{j^{U}}.
\end{equation}

When given all UAVs' locations during the $\tau_{0}$, we have \textbf{P2a} as follows:
\beqi\!\!\!\!\!\!\!\!\underline{\textbf{P2a:}}~~~~
&\min_{\textbf{I}_{K \times N}} &~F_{2a},\inum \label{1-2a_determine}\\
&{\rm s.t.} &~\frac{1}{\bar{D}_{i,j^{D}}}\geq\frac{I_{i,j^{D}}\rho}{P_{ut}},~\forall j^{D} \in  \mathcal{N},\inum\label{1-2a_constraint_EH1}\\
&&I_{i,j^{D}}\in\{0,1\},~\forall j^{D} \in  \mathcal{N},\inum\label{1-2a_constraint_EH2} \inum\label{1-2a_constraint_I}\\
&&\sum_{j^{D}=1}^{N}I_{i,j^{D}}\geq1, \inum\label{1-2a_constraint_sumI}
\eeqi
Here, \textbf{P2a} is a linear 0-1 fractional programming problem. Although we can use Isbell-Marlow procedure or Charnes-Cooper procedure transform it into the linear integer problems which can be solved by a cutting plane algorithm. However, the cutting plane algorithm can be inefficient for potentially high number of IoT devices in large-scale IoT. Consider $C(\pmb{x})=a_{0}+\sum_{j^{D}=1}^{N}a_{j^{D}}x_{j^{D}}$ donates the cost function, and $V(\pmb{x})=b_{0}+\sum_{j^{D}=1}^{N}b_{j^{D}}x_{j^{D}}$ donates the value function, a 0-1 fractional programming problem is as follows\cite{YOU20091879}:
\begin{equation}
  \min_{\pmb{x}}~~r=\frac{C(\pmb{x})}{V(\pmb{x})}, x_{j^{D}}\in\{0,1\}.
\end{equation}

For \textbf{P2a}, if $\varpi<1$, $a_{0}=b_{0}=\Phi_{i}[1+\Theta_{i}^{1}(k-1)]$, $a_{j^{D}}=-\tau_{1}^*/\bar{D}_{i,j^{D}}(\resizebox{0.095\hsize}{!}{$d_{i,j^{D}},\theta_{i,j^{D}}$})$ and $b_{j^{D}}=\tau_{0}^*/\bar{D}_{i,j^{D}}(\resizebox{0.095\hsize}{!}{$d_{i,j^{D}},\theta_{i,j^{D}}$})$. Otherwise, $a_{0}=1$, $b_{0}=\Phi_{i}[1+\Theta_{i}^{1}(k-1)]$, $a_{j^{D}}=0$ and $b_{j^{D}}=\Gamma_{i}/\bar{D}_{i,j^{D}}(\resizebox{0.095\hsize}{!}{$d_{i,j^{D}},\theta_{i,j^{D}}$})$. For generating a sequence of parameters converging to optimal $r^{*}$, there are various methods, such as a binary search method and a Dinkelbach's algorithm \cite{matsui92ananalysis}. It is shown for the nonlinear fractional programming problems that the convergence rate of the binary search method is linear, whereas Dinkelbach's algorithm converges superlinearly \cite{doi:10.1287/mnsc.22.8.868,YOU20091879}. Besides, compared to the conventional 0-1 fractional programming problem, \textbf{P2a} has an additional constraint in \eqref{1-2a_constraint_EH1} which results from the DL energy harvesting threshold constraint. This constraint indicates that device $i$ cannot be assigned to UAV $j^{D}$ if $P_{ut}/\bar{D}_{i,j^{D}}<\rho$. Therefore, in the 0-1 fractional programming problem, we can consider $a_{j^{D}}=+\infty$ to avoid assigning device $i$ to UAV $j^{D}$ when $P_{ut}/\bar{D}_{i,j^{D}}<\rho$ which implies the constraint in \eqref{1-2a_constraint_EH1} is violated. In this study, compared to a binary search method with a time complexity of $\mathcal{O}\left(KNB_{1}B_{2}\right)$ where $B_{1}=\max_{\forall i \in A_{j^{U}}^{k},~\forall k \in L_{j^{U}}}\{\max_{\forall j^{D}\in \mathcal{N}}  a_{j^{D}},1\}$ and $B_{2}=\max_{\forall i \in A_{j^{U}}^{k},~\forall k \in L_{j^{U}}}\{\max_{j^{D}\in \mathcal{N}} b_{j^{D}},1\}$, we employ the Dinkelbach's algorithm of 0-1 fractional programming to solve \textbf{P2a} with time complexity $\mathcal{O}\left(KNB_{3}\right)$, where $B_{3}=\max_{\forall i \in A_{j^{U}}^{k},~\forall k \in L_{j^{U}}}\{\max_{j^{D}\in \mathcal{N}} a_{j^{D}},\max_{j^{D}\in \mathcal{N}} b_{j^{D}},1\}$. Another major advantage of using Dinkelbach's algorithm is that no linear integer solver is required, and the memory usage during the computational process tends to be rather small compared with using cutting plane method or binary search method, especially for large-scale linear 0-1 fractional programming problems.

\subsubsection{3D Position Acquisition Step}\label{Sec_DL_Position}
After obtaining assignment matrix $\textbf{I}_{K \times N}$ to UAVs from \textbf{P2a}, we can then also obtain the set of devices served by UAV $j^{D}$, which is defined as $\mathcal{B}_{j^{D}}\triangleq\{k|I_{k,j^{D}}\!\!=\!\!1,\forall k \!\!\in\!\!  \mathcal{K}\}, \forall j^{D} \!\!\in \!\! \mathcal{N}$. For each UAV $j^{D}$, $F_{2}$ can be equivalently reformulated as follows:
\begin{equation}\label{F1-2b}
F_{2b}=\left\{
\begin{aligned}
&\sum_{k=1}^{L_{j^{U}}}\sum_{i=1}^{B_{j^{D}}^{k}}\frac{1}{\Phi_{i}[1+\Theta_{i}^{1}(k-1)]\bar{D}_{i,j^{D}}(\resizebox{0.095\hsize}{!}{$d_{i,j^{D}},\theta_{i,j^{D}}$}) +  \tau_{0}^*},~~\varpi<1,\\
&\sum_{k=1}^{L_{j^{U}}}\sum_{i=1}^{B_{j^{D}}^{k}} \frac{\Gamma_{i}/[1+\Theta_{i}^{1}(k-1)]}{[1+\Theta_{i}^{1}(k-1)]\bar{D}_{i,j^{D}}(\resizebox{0.095\hsize}{!}{$d_{i,j^{D}},\theta_{i,j^{D}}$})+\Gamma_{i}},~~\varpi\geq1,\
\end{aligned}
\right.
\end{equation}
where $B_{j^{D}}^{k}=|\mathcal{B}_{j^{D}}^{k}|$ is the number of devices served by UAV $j^{D}$ at epoch $k$.

It is noted that the objective functions \eqref{F1-2b} under conditions that $\varpi<1$ and $\varpi\geq1$ have the same structure, i.e.,
\begin{equation}
 F_{2b}=\sum_{k=1}^{L_{j^{U}}}\sum_{i=1}^{B_{j^{D}}^{k}}\frac{\alpha_{3,i,k}}{\alpha_{1,i,k}\bar{D}_{i,j^{D}}(\resizebox{0.095\hsize}{!}{$d_{i,j^{D}},\theta_{i,j^{D}}$}) +  \alpha_{2,i}},
\end{equation}
where $\alpha_{1,i,k}=\Phi_{i}[1+\Theta_{i}^{1}(k-1)]$, $\alpha_{2,i}=\tau_{0}^*$, and $\alpha_{3,i,k}=1$ if $\varpi<1$, and $\alpha_{1,i,k}=[1+\Theta_{i}^{1}(k-1)]$, $\alpha_{2,i}=\Gamma_{i}$, and $\alpha_{3,i,k}=\Gamma_{i}/[1+\Theta_{i}^{1}(k-1)]$ otherwise.

When assignment matrix $\textbf{I}_{K \times N}$ is given from \textbf{P2a}, for each UAV $j^{D}$, \textbf{P2} can be equivalently reformulated into the following problem:
\beqi\!\!\!\!\!\!\!\!\underline{\textbf{P2b:}}~~~~\label{P1-2b}
&\max_{\pmb{u}_{j^{D}}} &~F_{2b},\inum \label{1-2b_determine}\\
&{\rm s.t.} &~\frac{P_{ut}}{\bar{D}_{i,j^{D}}(x_{j^{D}},y_{j^{D}},h_{j^{D}})}\geq\rho,~\forall i \in  \mathcal{B}_{j^{D}}^{k},~\forall k \in \mathcal{L}_{j^{U}},~\forall j^{D} \in  \mathcal{N}.\inum\label{1-2b_constraint_wet1}
\eeqi

It is observed that the problem \eqref{P1-2b} is a sum-of-ratios programming problem. This is a non-convex and non-deterministic polynomial (NP)-hard problem \cite{stancu2012fractional}. The classical Dinkelbach transformation is usually used to solve the single-ratio fractional problems \cite{doi:10.1287/mnsc.13.7.492}. However, the Dinkelbach transformation cannot be easily generalized to the multiple-ratio fractional problems because the objective function of the transformed problem is not necessarily the same as the objective function of the original fractional problem. To tackle this issue, we employ Dinkelbach quadratic transformation which can guarantee a strong equivalence after the transformation by equivalently transforming the fractional term in the sum-of-ratios problem into a linear term \cite{8314727}. The problem is formulated as follows:
\beqi\label{1-2b1}
\!\!&\max_{\pmb{u}_{j^{D}},\pmb{\varsigma}}&~\sum_{k=1}^{L_{j^{U}}}\sum_{i=1}^{B_{j^{D}}^{k}}2\varsigma_{i}\sqrt{\alpha_{3,i,k}}\!-\!\varsigma_{i}^{2}\left(\alpha_{1,i,k}\kappa_{0}^{2}d_{i,j^{D}}^{2}[\mu^{\rm{LoS}}Pr_{i,j^{D}}^{\rm{LoS}}\!+\!\mu^{\rm{NLoS}}Pr_{i,j^{U}}^{\rm{NLoS}}]\!+\!  \alpha_{2,i}\right), \forall j^{D}\! \in \! \mathcal{N},\inum \label{1-2b1_determine}\\
&{\rm s.t.} 
&~d_{i,j^{D}}^{2}[\mu_{\rm{LoS}}Pr_{i,j^{D}}^{\rm{LoS}}+\mu_{\rm{NLoS}}Pr_{i,j^{D}}^{\rm{NLoS}}]\leq \frac{P_{ut}}{\kappa_{0}^{2}\rho},~\forall i \in  \mathcal{B}_{j^{D}}^{k},~\forall k \in \mathcal{L}_{j^{U}},~\forall j^{D} \in  \mathcal{N},\inum
\eeqi
where $\pmb{\varsigma}\triangleq\{\varsigma_{1},\varsigma_{2},\cdots,\varsigma_{B_{j^{D}}}\}$ is introduced as a slack vector.

In the $(m+1)$th iteration, given the slack variables $\pmb{\varsigma}^{(m)}$ in the $m$th iteration, the optimization problem is non-convex jointly over $(x_{j^{D}}, y_{j^{D}}, h_{j^{D}})$ because $Pr_{i,j^{D}}^{\rm{LoS}}$ is highly nonlinear, which makes the problem \eqref{1-2b1} difficult to handle for the purpose of UAV location optimization. However, given any altitude $h_{j^{D}}$, $Pr_{i,j^{D}}^{\rm{LoS}}$ is a decreasing function of $d_{i,j^{D}}$ and $\mu^{\mathrm{LoS}}Pr_{i,j}^{\mathrm{LoS}}\!+\!\mu^{\mathrm{NLoS}}Pr_{i,j}^{\mathrm{NLoS}}$ is an increasing function of $d_{i,j^{D}}$ since $\mu^{\mathrm{LoS}}<\mu^{\mathrm{NLoS}}$. Thus, we can reformulate this problem as a convex optimization problem.

\begin{figure}[t]
\centerline{\includegraphics[width = 8cm]{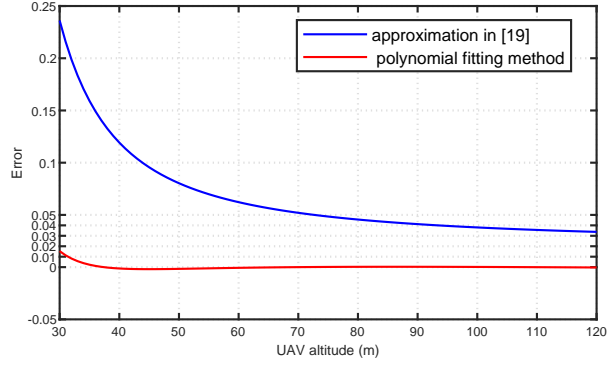}}
\vspace{-0.2cm}
\caption{Error from the objective function approximation.}
\vspace{-0.5cm}
\label{bias}
\end{figure}

Now, we consider $F(\resizebox{0.032\hsize}{!}{$d_{i,j}$})\!=\!d_{i,j}^{2}[\mu^{\mathrm{LoS}}Pr_{i,j}^{\mathrm{LoS}}\!+\!\mu^{\mathrm{NLoS}}Pr_{i,j}^{\mathrm{NLoS}}]$ that is used in problem \eqref{1-2b1}. Although $F(\resizebox{0.032\hsize}{!}{$d_{i,j}$})$ is approximated by a convex quadratic function in \cite{8038869}, a large bias exists owing to the approximation if the fixed altitude is small. Therefore, in pursuit of a more accurate result, we approximate it as a convex quadratic function by using a polynomial fitting method as follows:
\begin{equation}
  F(d_{i,j})\!\approx \!\left(k_{1}d_{i,j}\!+\!k_{2}\right)^{2}\!+\!k_{3},
\end{equation}
where $k_{1}$, $k_{2}$ and $k_{3}$ are altitude dependent coefficients. Fig. \ref{bias} shows the error of the objective function (29) caused by the quadratic approximation. As we can see from Fig. \ref{bias}, which is obtained based on the parameters in Table \ref{tabel_SS}, the error is less than $2\%$ for different UAVs’ altitudes.

%

Noting that $F(\resizebox{0.048\hsize}{!}{$d_{i,j^{D}}$})$ is a monotonically increasing function of feasible $d_{i,j^{D}}$, \textbf{P2b} can be reformulated for arbitrarily given $h_{j^{D}}$ as follows:
\beqi\underline{\textbf{P2b-1:}}\label{P1-2b-1}
&\max_{x_{j^{D}}^{(m+1)},y_{j^{D}}^{(m+1)}} &\sum_{k=1}^{L_{j^{U}}}\sum_{i=1}^{B_{j^{D}}^{k}}(2\varsigma_{i}\sqrt{\alpha_{3,i,k}}-\varsigma_{i}^{2}\alpha_{2,i})-\varsigma_{i}^{2}\alpha_{1,i,k}\kappa_{0}^{2}\left[\left(\!k_{1}\|\pmb{u}_{j^{D}}\!\!-\!\!\pmb{s}_{i}\|_{2}\!+\!k_{2}\!\right)^{2}\!+\!k_{3}\right],\inum \label{1-2b-1_determine}\\
&{\rm s.t.} &~\|\pmb{u}_{j^{D}}-\pmb{s}_{i}\|_{2}^{2}\leq (d_{0}^{m})^{2},~\forall i \in  \mathcal{B}_{j^{D}}^{k},~\forall k \in \mathcal{L}_{j^{U}},~\forall j^{D} \in  \mathcal{N},\inum\label{1-2b-1_constraint_wet1}
\eeqi
where $d_{0}^{m}=\mathrm{Root}\Big(F(\resizebox{0.048\hsize}{!}{$d_{i,j^{D}}$})=P_{ut}/(\kappa_{0}^{2}\rho)\Big)$.

As we can see from \textbf{P2b-1}, $\varsigma_{i}^{2}\alpha_{1,i,k}\kappa_{0}^{2}\left[\left(\!k_{1}\|\pmb{u}_{j^{D}}\!\!-\!\!\pmb{s}_{i}\|_{2}\!+\!k_{2}\!\right)^{2}\!+\!k_{3}\right]$ is a strictly convex function of $\pmb{u}_{j^{D}}\!-\!s_{i}$, because norm is a convex function and the power function $f(x)=x^{a}$ is a convex and non-decreasing function if $a\geq1$. Thus, according to \cite{boyd2004convex}, problem \eqref{P1-2b-1} is concave, which can be solved by using the off-the-shelf solvers, e.g., CVX. After we get the suboptimal 2D coordinates of UAVs, $(\pmb{l}_{j^{D}}^{*})^{(m\!+\!1)}=((x_{j^{D}}^{*})^{(m\!+\!1)},(y_{j^{D}}^{*})^{(m\!+\!1)})$ in the $m$th iteration, the suboptimal altitudes of UAVs can be obtained by solving the following problem:
\begin{equation}\label{P2b-2}
\begin{split}
&\underline{\textbf{P2b-2:}}~~(h_{j^{D}}^*)^{(m+1)}=\arg\max_{h_{j^{D}}} \sum_{k=1}^{L_{j^{U}}}\sum_{i=1}^{B_{j^{D}}^{k}}\Big((2\varsigma_{i}\sqrt{\alpha_{3,i,k}}-\varsigma_{i}^{2}\alpha_{2,i})-\\
\!\!\!\!\!\!\!\!\!\!\!\!\!\!\!\!\!\!\!\!\!\!\!\!\!\!\!\!\!\!\!\!\!\!\!\!&\varsigma_{i}^{2}\alpha_{1,i,k}\kappa_{0}^{2}\|((\pmb{l}_{j^{D}}^{*})^{(m\!+\!1)},h_{j^{D}})\!-\!\pmb{s}_{i}\|_{2}^{2}\Big[\frac{\mu^{\rm{LoS}}\!-\!\mu^{\rm{NLoS}}}{1\!\!+\!\!\beta \exp(\!-\!\psi[\frac{180}{\pi}\arcsin\Big(\frac{h_{j^{U}}}{\sqrt{\|((\pmb{l}_{j^{D}}^{*})^{(m\!+\!1)},h_{j^{D}})\!-\!\pmb{s}_{i}\|_{2}^{2}}}\Big)\!\!-\!\!\beta])}\!+\!\mu^{\rm{NLoS}}\Big]\Big),
\end{split}
\end{equation}
where \eqref{P2b-2} is still a highly nonlinear function, and its convexity can be judged by the second-order Hessian matrix. Here, through the numerical simulation with the parameters in Table \ref{tabel_SS}, it is verified \eqref{P2b-2} is a one-dimensional quasi-concave function. The suboptimal altitude of the UAV is then obtained via one-dimensional search over a feasible range of altitudes.

After $(\pmb{u}_{j^{D}}^{*})^{(m+1)}=((x_{j^{D}}^{*})^{(m+1)},(y_{j^{D}}^{*})^{(m+1)},(h_{j^{D}}^{*})^{(m+1)})$ in the $(m\!+\!1)$th iteration is obtained, an efficient gradient descent algorithm can be applied to obtain the optimal value of $\varsigma_{i}^{(m+1)}$ by letting the gradient of objective function \eqref{1-2b1_determine} respect to $\varsigma_{i}$ equals to a zero. The resultant $\varsigma_{i}^{(m+1)}$ is as follows:
\begin{equation}\label{varsigma_update}
  \varsigma_{i}^{(m+1)}=\frac{\sqrt{\alpha_{3,i,k}}}{\alpha_{1,i,k}\kappa_{0}^{2}d_{i,j^{D}}^{2}[\mu^{\rm{LoS}}Pr_{i,j^{D}}^{\rm{LoS}}+\mu^{\rm{NLoS}}Pr_{i,j^{U}}^{\rm{NLoS}}]+  \alpha_{2,i}}.
\end{equation}

Note that, \eqref{P1-2b-1}, \eqref{P2b-2}, and \eqref{varsigma_update} are guaranteed to achieve a stationary point of the concave-convex fractional programming problems with a nondecreasing sum-of-functions-of-ratio value in each iteration. Furthermore, Dinkelbach quadratic transformation allows the algorithm to explore the solution space almost fully \cite{8314727}.


\subsection{UL assignment and 3D location optimization}\label{Sec_UL}
For the given optimal time allocation, UL scheduling, DL user association, and UAV locations, we optimize the UL assignment and 3D locations of UAVs. Based on Theorem \ref{theorm-1-time}, the optimization problem \textbf{P1} under condition that $\varpi<1$ can be reformulated as follows:
\begin{align}
  &\max_{\textbf{L}^{U},\pmb{\mathcal{C}},\pmb{\mathcal{Z}}^{U}} ~\sum_{j^{U}=1}^{N}\sum_{i=1}^{C_{j^{U}}}\sum_{k=1}^{L_{j^{U}}} \frac{\tau_{1}^{*}}{L_{j^{U}}}R_{i,j^{U}}^{k} \Longleftrightarrow\max_{\textbf{L}^{U},\pmb{\mathcal{C}},\pmb{\mathcal{Z}}^{U}} ~\sum_{j^{U}=1}^{N}\sum_{i=1}^{C_{j^{U}}}\sum_{k=1}^{L_{j^{U}}} \frac{s_{i,k}^{*}\Theta_{i}^{0}L_{j^{U}}}{\Theta_{i}^{0}L_{j^{U}}\tau_{0}^{*}+[1+\Theta_{i}^{1}(k-1)]\tau_{1}^{*} }, \\
  &\Longleftrightarrow\max_{\textbf{L}^{U},\pmb{\mathcal{C}},\pmb{\mathcal{Z}}^{U}} ~\sum_{j^{U}=1}^{N}\sum_{i=1}^{C_{j^{U}}}\sum_{k=1}^{L_{j^{U}}}\frac{s_{i,k}^{*}}{\Lambda_{i}/g_{i,j^{U}}(\resizebox{0.08\hsize}{!}{$d_{i,j^{U}},\theta_{i,j^{U}}$}) + \varepsilon_{i}\Lambda_{i}(k-1)\sum_{j^{U}\in\mathcal{Z}_{i}^{U}}g_{i,j^{U}}(\resizebox{0.095\hsize}{!}{$d_{i,j^{U}},\theta_{i,j^{U}}$})+ \tau_{0}^*},\notag
\end{align}
where $\Lambda_{i}=\frac{\tau_{1}^{*}}{\varepsilon_{i}L_{j^{U}}\sum_{j^{D}\in\mathcal{Z}_{i}^{D}}g_{i,j^{D}}(\resizebox{.01\hsize}{!}{\textit{d}}_{i,j^{D}})}$.

On the other hand, the optimization problem \textbf{P1} under condition that $\varpi\geq1$ can be reformulated as follows:
\begin{align}
  &\max_{\textbf{L}^{U},\pmb{\mathcal{C}},\pmb{\mathcal{Z}}^{U}} \sum_{j^{U}=1}^{N}\sum_{i=1}^{C_{j^{U}}}\sum_{k=1}^{L_{j^{U}}} \frac{\tau_{1}^{*}}{L_{j^{U}}}R_{i,j^{U}}^{k} \Longleftrightarrow\min_{\textbf{L}^{U},\pmb{\mathcal{C}},\pmb{\mathcal{Z}}^{U}} \sum_{j^{U}=1}^{N}\sum_{i=1}^{C_{j^{U}}}\sum_{k=1}^{L_{j^{U}}} \frac{s_{i,k}^{*}\Theta_{m}^{0}}{\Theta_{i}^{0}[\gamma\!-\!\Theta_{m}^{1}(n\!-\!1)]\!+\![1\!+\!\Theta_{i}^{1}(k\!-\!1)]\Theta_{m}^{0}},\\
  &\!\!\Longleftrightarrow\!\!\max_{\textbf{L}^{U},\pmb{\mathcal{C}},\pmb{\mathcal{Z}}^{U}}\!\! \sum_{j^{U}=1}^{N}\sum_{i=1}^{C_{j^{U}}}\sum_{k=1}^{L_{j^{U}}} \notag \frac{s_{i,k}^{*}\left(\Omega_{i}\!+\!\varepsilon_{i}(k\!-\!1)\sum_{n^{U}\in\mathcal{Z}_{i}^{U}}g_{i,n^{U}}(\resizebox{0.095\hsize}{!}{$d_{i,n^{U}},\theta_{i,n^{U}}$})\right)}{\Omega_{i}\!+\!\varepsilon_{i}(k\!-\!1)\sum_{n^{U}\in\mathcal{Z}_{i}^{U}}g_{i,n^{U}}(\resizebox{0.095\hsize}{!}{$d_{i,n^{U}},\theta_{i,n^{U}}$})\!+\!1/g_{i,j^{U}}(\resizebox{0.095\hsize}{!}{$d_{i,j^{U}},\theta_{i,j^{U}}$})}\!,
\end{align}
where $\Omega_{i}=\frac{\varepsilon_{i}\sum_{j^{D}\in\mathcal{Z}_{i}^{D}}g_{i,j^{D}}(\resizebox{.08\hsize}{!}{$\textit{d}_{i,j^{D}},\theta_{i,j^{D}}$})[\gamma-\Theta_{m}^{1}(n-1)]}{\Theta_{m}^{0}}$.

In this section, we thus mainly discuss on the optimization of the following function:
\begin{equation}\label{F3}
F_{3}\!\!=\!\!\left\{
\begin{aligned}
&\!\!\sum_{j^{U}=1}^{N}\!\sum_{i=1}^{C_{j^{U}}}\!\sum_{k=1}^{L_{j^{U}}}\!\!\frac{s_{i,k}^{*}}{\Lambda_{i}/g_{i,j^{U}}(\resizebox{0.08\hsize}{!}{$d_{i,j^{U}},\theta_{i,j^{U}}$}) + \varepsilon_{i}\Lambda_{i}(k-1)\sum_{n^{U}\in\mathcal{Z}_{i}^{U}}g_{i,n^{U}}(\resizebox{0.095\hsize}{!}{$d_{i,n^{U}},\theta_{i,n^{U}}$})+ \tau_{0}^*},~\varpi<1,\\
&\!\!\sum_{j^{U}=1}^{N}\!\sum_{i=1}^{C_{j^{U}}}\!\sum_{k=1}^{L_{j^{U}}} \!\!\frac{s_{i,k}^{*}\!\!\left(\Omega_{i}\!+\!\varepsilon_{i}(k\!-\!1)\sum_{n^{U}\in\mathcal{Z}_{i}^{U}}g_{i,n^{U}}(\resizebox{0.095\hsize}{!}{$d_{i,n^{U}},\theta_{i,n^{U}}$})\right)}{\Omega_{i}\!+\!\varepsilon_{i}(k\!-\!1)\sum_{n^{U}\in\mathcal{Z}_{i}^{U}}g_{i,n^{U}}(\resizebox{0.095\hsize}{!}{$d_{i,n^{U}},\theta_{i,n^{U}}$})\!+\!1/g_{i,j^{U}}(\resizebox{0.095\hsize}{!}{$d_{i,j^{U}},\theta_{i,j^{U}}$})},~o.w.\
\end{aligned}
\right.
\end{equation}

Given the optimal time allocation, UL scheduling, DL locations of UAVs, and DL association between UAVs and devices, we optimize jointly the UL assignment and 3D locations of UAVs by solving the following problem:
\beqi\!\!\!\!\!\!\!\!\underline{\textbf{P3:}}~~~~
&\max_{\textbf{L}^{U},\pmb{\mathcal{C}},\pmb{\mathcal{Z}}^{U}} &~F_{3}\inum \label{P3_determine}\\
&{\rm s.t.} &~\eqref{EH_constraint2},\eqref{DL_constraint2},\eqref{SNR_constraint},~\eqref{UL_constraint1} ,~\mbox{and}~ \eqref{UL_constraint2}.\notag
\eeqi

Similar to the DL optimization process, directly solving \textbf{P3} is challenging, because the UAVs' locations and device association are mutually coupled. Thus, we decompose \textbf{P3} into two subproblems. In the first subproblem, given the fixed locations of UAVs, the UL devices associations are optimized. In the second subproblem, given the optimal device association obtained from the first subproblem, the suboptimal UL 3D locations of UAVs are designed to maximize the sum throughput of the devices.

Here, we note that satisfying the SNR requirement \eqref{SNR_constraint} of each device significantly depends on the distance and the elevation angle between the device and its serving UAV.
\subsubsection{UL Assignment Step}
In UL phase, UAVs not only provide energy but also gather information. Denote an allocation matrix by $\textbf{A}_{K \times N}$. Here, if the $i$th device is assigned to the $j^U$th UAV, the $(i,j^U)$th element $a_{i,j^U}$ of $\textbf{A}_{K \times N}$ is one, and otherwise a zero. The set of devices served by UAV $j^{U}$ is then defined as follows:
\begin{equation}
  \mathcal{C}_{j^{U}}\triangleq\{k|a_{k,j^{U}}=1,~\forall k \in  \mathcal{K}\}, ~\forall j^{U} \in  \mathcal{N}.
\end{equation}

Similar to \textbf{P2a}, we introduce a binary variable, $b_{i,j^{U}}$, where $b_{i,j^{U}}=1$ if device $i$ can harvest energy from radio signals from UAV $j^{U}$, and $b_{i,j^{U}}=0$ otherwise. The corresponding allocation matrix is denoted by $\textbf{B}_{K \times N}$. The set of UAVs that covers device $i$ is then defined as follows:
\begin{equation}
  \mathcal{Z}_{i}^{U}\triangleq\{n|b_{i,n}=1, ~\forall n \in  \mathcal{N}\},~\forall i \in  \mathcal{K}.
\end{equation}

Using $a_{i,j^{U}}$ and $b_{i,j^{U}}$, $F_{3}$ can be reformulated as follows:
\begin{equation}\label{F3a}
F_{3a}=\left\{
\begin{aligned}
&\frac{s_{i,k}^{*}a_{i,j^{U}}}{\Lambda_{i} \bar{D}_{i,j^{U}}(\resizebox{0.095\hsize}{!}{$d_{i,j^{U}},\theta_{i,j^{U}}$}) \!+\! \varepsilon_{i}\Lambda_{i}(k\!-\!1)\sum_{n^{U}=1}^{N}b_{i,n^{U}}/\bar{D}_{i,n^{U}}(\resizebox{0.095\hsize}{!}{$d_{i,n^{U}},\theta_{i,n^{U}}$})\!+\! \tau_{0}^*},~\varpi<1,\\
&\frac{s_{i,k}^{*}a_{i,j^{U}}\left(\Omega_{i}+\varepsilon_{i}(k-1)\sum_{n^{U}=1}^{N}b_{i,n^{U}}/\bar{D}_{i,n^{U}}(\resizebox{0.095\hsize}{!}{$d_{i,n^{U}},\theta_{i,n^{U}}$})\right)}{\Omega_{i}\!+\!\varepsilon_{i}(k\!-\!1)\sum_{n^{U}=1}^{N}b_{i,n^{U}}/\bar{D}_{i,n^{U}}(\resizebox{0.095\hsize}{!}{$d_{i,n^{U}},\theta_{i,n^{U}}$})\!+\!\bar{D}_{i,j^{U}}(\resizebox{0.095\hsize}{!}{$d_{i,j^{U}},\theta_{i,j^{U}}$})},~\varpi\geq1.\
\end{aligned}
\!\!\right.,\forall i \in  \mathcal{K}.
\end{equation}

There exist the DL energy service constraints \eqref{DL_constraint2} and the UL information service constraints \eqref{UL_constraint1} and \eqref{UL_constraint2}. Since different constraints in DL energy transfer and UL information transmission, it is difficult to directly obtain the optimal UAV-device association in UL. Therefore, we use an alternating optimization method. We first optimize the UAV-device association variable $b_{i,j^{U}}$ for energy transmission with the fixed information transmission device-UAV association variable $a_{i,j^{U}}$ and locations of UAVs, which is then be formulated as follows:
\beqi\!\!\!\!\!\!\!\!\underline{\textbf{P3a-1:}}~~~~
&\max_{\textbf{B}_{K \times N}} &~F_{3a},\inum \label{1b_determine}\\
&{\rm s.t.} &~\frac{1}{\bar{D}_{i,j^{U}}}\geq\frac{b_{i,n^{U}}\rho}{P_{ut}},~\forall n^{U} \in  \mathcal{N},\inum\label{1b_constraint_wet1}\\
&&b_{i,n^{U}}\in\{0,1\},~\forall n^{U} \in  \mathcal{N},\inum \label{1b_constraint_tau0}\\
&&\sum_{n^{U}=1}^{N}b_{i,n^{U}}\geq1. \inum\label{1b_constraint_tau0}
\eeqi

Because binary variable $b_{i,n^{U}}$ in \textbf{P3a-1} is located in the denominator, \textbf{P3a-1} is a binary fractional programming problem, which can also be solved by using the same method of \textbf{P2a}. Using obtaining assignment matrix $\textbf{B}_{K \times N}$ to UAVs from \textbf{P3a-1}, we can then obtain $\textbf{A}_{K \times N}$ by solving following problem:\vspace{-0.2cm}
\beqi\!\!\!\!\!\!\!\!\underline{\textbf{P3a-2:}}~~~~
&\max_{\textbf{A}_{K \times N}} &~\sum_{j^{U}=1}^{N}\sum_{i=1}^{K}\sum_{k=1}^{L_{j^{U}}} F_{3a},\inum \label{1b_determine}\\
&{\rm s.t.} &~\sum_{j^{U}=1}^{N}a_{i,j^{U}}=1,~\forall i \in \mathcal{K}, \inum\label{1b_constraint_tau0}\\
&&\frac{a_{i,j^{U}}\tau_{1}s_{i,k}^{*}\gamma\bar{D}_{i,j^{U}}}{\varepsilon_{i}}\leq \chi_{i,j^{U},k}, ~\forall i \in  \mathcal{K},~\forall j^{U} \in  \mathcal{N}, \inum\label{1b_constraint_tau0}
\eeqi
where $\chi_{i,j^{U},k}\triangleq L_{j^{U}}\tau_{0}\sum_{j^{D}\in\mathcal{Z}_{i}^{D}}g_{i,j^{D}}+(k-1)\tau_{1}\sum_{n^{U}=1}^{N}\frac{b_{i,n^{U}}}{\bar{D}_{i,n^{U}}}$. Since the binary variable $a_{i,j^{U}}$ is involved in the numerator only, \textbf{P3a-2} is evidently an linear integer programming problem, and we can solve it by using a standard linear integer algorithm, i.e., a cutting plane method. However, the cutting plane method can be inefficient for potentially high number of IoT devices in large-scale IoT. Similar to \textbf{P2a}, after setting the cost value to be $+\infty$ to avoid the constraint in \eqref{1b_constraint_tau0} being violated, we can transform \textbf{P3a-2} to a classical assignment problem, which can be solved by using a Hungarian method with a complexity order of $\mathcal{O}((KN)^{3})$\cite{doi:10.1002/nav.20053}.

\subsubsection{3D Position Acquisition Step}

It is worth noting that a device may harvest energy from the signals emitted by more than one UAV in the DL energy harvest scenario, owing to the broadcast nature of the wireless medium. However, a device can only contact one UAV in the UL data communications. Therefore, the DL device set $\mathcal{B}$ and UL device set $\mathcal{C}$ served by different UAVs have the following relationship: $\mathcal{B}_{m_{0}}\cap\mathcal{B}_{n_{0}}\neq\emptyset$ and $ \mathcal{C}_{m_{1}}\cap\mathcal{C}_{n_{1}}=\emptyset, \exists m_{0},n_{0}\in  \mathcal{N}, \forall m_{1},n_{1} \in  \mathcal{N}$.

For each UAV $j^{U}$, we further reformulate $F_{3}$ as follows:
\begin{equation}\label{F1-3b_new}
F_{3b}\!=\!\left\{
\begin{aligned}
&\sum_{k=1}^{L_{j^{U}}}\sum_{i=1}^{C_{j^{U}}^{k}}\frac{v(\resizebox{0.035\hsize}{!}{$b_{i,j^{U}}$})}{\Lambda_{i} \bar{D}_{i,j^{U}}(\resizebox{0.095\hsize}{!}{$d_{i,j^{U}},\theta_{i,j^{U}}$})v(\resizebox{0.035\hsize}{!}{$b_{i,j^{U}}$}) \!+\! (\Lambda_{i}\iota_{i,k}+\tau_{0}^{*})v(\resizebox{0.035\hsize}{!}{$b_{i,j^{U}}$})\!+\!\varepsilon_{i}\Lambda_{i}(k\!-\!1)b_{i,j^{U}}},~~\varpi<1,\\
&\sum_{k=1}^{L_{j^{U}}}\sum_{i=1}^{C_{j^{U}}^{k}} \frac{(\iota_{i,k}+\Omega_{i})v(\resizebox{0.035\hsize}{!}{$b_{i,j^{U}}$})\!+\!\varepsilon_{i}(k\!-\!1)b_{i,j^{U}}}{\bar{D}_{i,j^{U}}(\resizebox{0.095\hsize}{!}{$d_{i,j^{U}},\theta_{i,j^{U}}$})v(\resizebox{0.035\hsize}{!}{$b_{i,j^{U}}$})\!+\! (\iota_{i,k}+\Omega_{i})v(\resizebox{0.035\hsize}{!}{$b_{i,j^{U}}$})\!+\!\varepsilon_{i}(k\!-\!1)b_{i,j^{U}}},~~\varpi\geq1,
\end{aligned}
\right.
\end{equation}
where $\iota_{i,k}\!\triangleq\!\varepsilon_{i}(k\!-\!1)\sum_{n^{U}=1,n^{U}\neq j^{U}}^{N}b_{i,n^{U}}/\bar{D}_{i,n^{U}}(\resizebox{0.095\hsize}{!}{$d_{i,j^{U}},\theta_{i,n^{U}}$})$, $v(\resizebox{0.035\hsize}{!}{$b_{i,j^{U}}$})\!\triangleq\!\left[(\bar{D}_{i,j^{U}}\!-\!1)b_{i,j^{U}}\!+\!1\right]$, and $C_{j^{U}}^{k}=|\mathcal{C}_{j^{U}}^{k}|$ is the number of devices served by UAV $j^{U}$ at epoch $k$.

It should be also noted that the objective functions \eqref{F1-3b_new} under conditions that $\varpi<1$ and $\varpi\geq1$ have the same structure as
\begin{equation}\label{F3b}
 F_{3b}=\sum_{k=1}^{L_{j^{U}}}\sum_{i=1}^{C_{j^{U}}^{k}}\frac{\varphi_{1,i,k} v(\resizebox{0.035\hsize}{!}{$b_{i,j^{U}}$})+\varphi_{2,i,k}b_{i,j^{U}} }{\varrho_{1,i}\bar{D}_{i,j^{U}}(\resizebox{0.095\hsize}{!}{$d_{i,j^{U}},\theta_{i,j^{U}}$})v(\resizebox{0.035\hsize}{!}{$b_{i,j^{U}}$}) +  \varrho_{2,i,k}v(\resizebox{0.035\hsize}{!}{$b_{i,j^{U}}$})+\varrho_{3,i,k}b_{i,j^{U}}},
\end{equation}
where $\varphi_{1,i,k}=1$, $\varphi_{2,i,k}=0$, $\varrho_{1,i}=\Lambda_{i}$, $\varrho_{2,i,k}=\Lambda_{i}\iota_{i,k}+\tau_{0}^{*}$, and $\varrho_{3,i,k}=\varepsilon_{i}\Lambda_{i}(k-1)$ if $\varpi<1$, whereas $\varphi_{1,i,k}=\iota_{i,k}+\Omega_{i}$, $\varphi_{2,i,k}=\varepsilon_{i}(k-1)$, $\varrho_{1,i}=1$, $\varrho_{2,i,k}=\varphi_{1,i,k}$, and $\varrho_{3,i,k}=\varphi_{2,i,k}$ if $\varpi\geq1$.

For the sum-of-ratios objective function \eqref{F3b}, using polynomial fitting and Dinkelbach quadratic transformation in Section \ref{Sec_DL_Position}, we formulate UL UAV placement optimization problem as follows:
\beqi\label{P1-3b}
&\!\!\!\!\!\!\!\max_{\pmb{u}_{j^{U}},\pmb{\xi}}\!\!&\sum_{k=1}^{L_{j^{U}}}\sum_{i=1}^{C_{j^{U}}^{k}}2\xi_{i}\sqrt{\varphi_{1,i,k} v(\resizebox{0.035\hsize}{!}{$b_{i,j^{U}}$})\!\!+\!\!\varphi_{2,i,k}b_{i,j^{U}}}\!\!-\!\!\xi_{i}^{2}(\varrho_{1,i}\bar{D}_{i,j^{U}}(\pmb{u}_{j^{U}})v(\resizebox{0.035\hsize}{!}{$b_{i,j^{U}}$}) \!\!+\!\!  \varrho_{2,i,k}v(\resizebox{0.035\hsize}{!}{$b_{i,j^{U}}$})\!\!+\!\!\varrho_{3,i,k}b_{i,j^{U}}\!), \inum \label{1-3b_determine}\\
&{\rm s.t.} 
&~d_{i,j^{U}}^{2}[\mu^{\rm{LoS}}Pr_{i,j^{U}}^{\rm{LoS}}+\mu^{\rm{NLoS}}Pr_{i,j^{U}}^{\rm{NLoS}}]\leq\frac{P_{ut}}{\kappa_{0}^{2}\rho},~\forall i \in  \mathcal{C}_{j^{U}}^{k},~\forall k \in \mathcal{L}_{j^{U}},~\forall j^{U} \in  \mathcal{N},\inum\\
&&~d_{i,j^{U}}^{2}[\mu^{\rm{LoS}}Pr_{i,j^{U}}^{\rm{LoS}}+\mu^{\rm{NLoS}}Pr_{i,j^{U}}^{\rm{NLoS}}]\leq \frac{\varepsilon_{i}}{\gamma\kappa_{0}^{2}}\Big(L_{j^{U}}\frac{\tau_{0}^{*}}{\tau_{1}^{*}}\sum_{j^{D}\in\mathcal{Z}_{i}^{D}}g_{i,j^{D}}+(k-1)\sum_{n^{U}=1}^{N}\frac{b_{i,n^{U}}}{\bar{D}_{i,n^{U}}}\Big),\inum
\eeqi
where $\pmb{\xi}\triangleq\{\xi_{1},\xi_{2},\cdots,\xi_{C_{j^{U}}}\}$ is introduced as a slack vector.

For convenience, we define the following function:
\begin{equation}\label{f_3b-1}
f_{i,j^{U}}^{3b-1}\!\!\triangleq\!\!\left\{
\begin{aligned}
&\!\!2\xi_{i}\sqrt{\varphi_{1,i,k}}-\xi_{i}^{2}\varrho_{2,i,k}-\xi_{i}^{2}\varrho_{1,i}\kappa_{0}^{2}\left[\left(\!k_{1}\|\pmb{u}_{j^{U}}\!\!-\!\!\pmb{s}_{i}\|_{2}\!+\!k_{2}\!\right)^{2}\!\!+\!\!k_{3}\right],~~b_{i,j^{U}}=0,\\
&\!\!U_{i,j^{U}}^{a}-U_{i,j^{U}}^{b},~~b_{i,j^{U}}=1.
\end{aligned}
\right.
\end{equation}

Here, for UAV placement, UAVs not only provide energy but also gather information in UL. Thus, the objective function \eqref{f_3b-1} has two states depending on whether this UAV only collects information from this device or it also transfers additional energy in the UL phase. When $b_{i,j^{U}}=0$, UAV $j^{U}$ only collects information from device $i$. Here, the objective function $f_{i,j^{U}}^{3b-1}$ is similar to the objective function \eqref{1-2b-1_determine} in the DL UAV placement optimization, where both are the concave functions. On the other hand, when $b_{i,j^{U}}=1$, UAV $j^{U}$ not only collects information from device $i$ but also transfers energy to device $i$ in previous epochs in UL phase. $U_{i,j^{U}}^{a}(\pmb{u}_{j^{U}})\triangleq c(\pmb{u}_{j^{U}})-\xi_{i}^{2}\varrho_{3,i,k}$ and $c(\pmb{u}_{j^{U}})\triangleq\xi_{i}\sqrt{2\varphi_{1,i,k}}\kappa_{0}\left(\!k_{1}\|\pmb{u}_{j^{U}}\!\!-\!\!\pmb{s}_{i}\|_{2}\!+\!k_{2}\!\right)+\xi_{i}\sqrt{2\varphi_{1,i,k}\kappa_{0}^{2}k_{3}+2\varphi_{2,i,k}}$ . $c(\pmb{u}_{j^{U}})$ is the lower bound of $2\xi_{i}\sqrt{\varphi_{1,i,k}\bar{D}_{i,j^{U}}(\pmb{u}_{j^{U}})\!\!+\!\!\varphi_{2,i,k}}$ from the fact that $\frac{a+b}{2}\leq\sqrt{\frac{a^{2}+b^{2}}{2}}$, and when $\|\pmb{u}_{j^{U}}\!\!-\!\!\pmb{s}_{i}\|_{2}=\frac{\sqrt{2\varphi_{1,i,k}k_{3}+2\varphi_{2,i,k}}}{k_{1}\sqrt{2\varphi_{1,i,k}}}-\frac{k_{2}}{k_{1}}$, $c(\pmb{u}_{j^{U}})$ is tight. $U_{i,j^{U}}^{b}(\pmb{u}_{j^{U}})\triangleq\xi_{i}^{2}\varrho_{1,i}\kappa_{0}^{4}\left(F(\|\pmb{u}_{j^{U}}\!\!-\!\!\pmb{s}_{i}\|_{2})\right)^{2}+\xi_{i}^{2}\varrho_{2,i,k}\kappa_{0}^{2}F(\|\pmb{u}_{j^{U}}\!-\!\pmb{s}_{i}\|_{2})$.  Since $U_{i,j^{U}}^{a}(\pmb{u}_{j^{U}})$ and $U_{i,j^{U}}^{b}(\pmb{u}_{j^{U}})$  are convex with respect to $(x_{j^{U}}, y_{j^{U}})$, $f_{i,j^{U}}^{3b-1}$ with $b_{i,j^{U}}=1$ has the difference-of-convex structure, which can be solved by the concave-convex procedure (CCCP) \cite{yuille2002concave}. Denoting $(x_{j^{U}}^{(n)}, y_{j^{U}}^{(n)}, h_{j^{U}}^{(n)})$ by $\pmb{u}_{j^{U}}^{(n)}$ at the fixed point at the $n$th iteration, the first-order Taylor series expansion of $U_{i,j^{U}}^{a}$ around $\pmb{u}_{j^{U}}^{(n)}$ can be expressed as follows:
\begin{equation}\label{tay}
U_{i,j^{U}}^{c}(\pmb{u}_{j^{U}}^{(n+1)})=U_{i,j^{U}}^{a}(\pmb{u}_{j^{U}}^{(n)})+\nabla\left(U_{i,j^{U}}^{a}(\pmb{u}_{j^{U}}^{(n)})\right)(\pmb{u}_{j^{U}}^{(n+1)}-\pmb{u}_{j^{U}}^{(n)}).
\end{equation}

By substituting \eqref{tay} into $f_{i,j^{U}}^{3b-1}$ in \eqref{f_3b-1} when $b_{i,j^{U}}=1$, we can obtain the lower bound of $f_{i,j^{U}}^{3b-1}$ when $b_{i,j^{U}}=1$ as follows:
\begin{equation}
f_{i,j^{U}}^{3b-1}=U_{i,j^{U}}^{c}-U_{i,j^{U}}^{b}.
\end{equation}

Noting that $U_{i,j^{U}}^{c}$ is an affine function with respect to $(x_{j^{U}}, y_{j^{U}})$ and $U_{i,j^{U}}^{b}$ is a convex function with respect to $(x_{j^{U}}, y_{j^{U}})$. Thus, the lower bound of $f_{i,j^{U}}^{3b-1}$ is concave over $(x_{j^{U}}, y_{j^{U}})$, which can be solved by using the standard convex optimization tools, e.g., CVX.

Thus, we can rewrite the optimization problem for given $h_{j^{U}}$ as follows:
\beqi\!\!\!\!\!\!\!\!\underline{\textbf{P3b-1:}}\label{P3b-1}~~~~
&\max_{x_{j^{U}}, y_{j^{U}}}&~\sum_{k=1}^{L_{j^{U}}}\sum_{i=1}^{C_{j^{U}}^{k}}f_{i,j^{U}}^{3b-1}, ~ \forall j^{U} \in  \mathcal{N},\inum \label{1-3b-1_determine}\\
&{\rm s.t.} 
&~\|\pmb{u}_{j^{U}}-\pmb{s}_{i}\|_{2}^{2}\leq (d_{i}^{m})^{2},~\forall i \in  \mathcal{C}_{j^{U}}^{k},~\forall k \in \mathcal{L}_{j^{U}},~\forall j^{U} \in  \mathcal{N},\inum
\eeqi
where $d_{i}^{m}\triangleq\min\left\{\mathrm{Root}\Big(F(\resizebox{0.04\hsize}{!}{$d_{i,j^{U}}$})=\frac{\varepsilon_{i}}{\gamma\kappa_{0}^{2}}\Big(L_{j^{U}}\frac{\tau_{0}^{*}}{\tau_{1}^{*}}\sum_{j^{D}\in\mathcal{Z}_{i}^{D}}g_{i,j^{D}}+(k-1)\sum_{n^{U}=1}^{N}\frac{b_{i,n^{U}}}{\bar{D}_{i,n^{U}}}\right)\!\Big),\mathrm{Root}\Big($\\$F(\resizebox{0.04\hsize}{!}{$d_{i,j^{U}}$})=\frac{P_{ut}}{\kappa_{0}^{2}\rho}\Big)\Big\}$.

Similar to \textbf{P3b-2}, we introduce the following one-dimensional function with $h_{j^{U}}$ as variable:
\begin{equation}
f_{i,j^{U}}^{3b-2}\!\!\triangleq\!\!\left\{
\begin{aligned}
&2\xi_{i}\sqrt{\varphi_{1,i,k}}\!\!-\!\!\xi_{i}^{2}\varrho_{2,i,k}\!\!-\!\!\xi_{i}^{2}\varrho_{1,i}\mathcal{D}\left(\|((\pmb{l}_{j^{U}}^{*})^{(n+1)},h_{j^{U}})\!\!-\!\!\pmb{s}_{i}\|_{2}^{2}\right),~~b_{i,j^{U}}\!\!=\!\!0,\\
&2\xi_{i}\sqrt{\varphi_{1,i,k} \mathcal{D}\!\left(\!\|((\pmb{l}_{j^{U}}^{*})^{(n+1)},h_{j^{U}})\!\!-\!\!\pmb{s}_{i}\|_{2}^{2}\!\right)\!\!+\!\!\varphi_{2,i,k}}\!\!-\!\!\xi_{i}^{2}\Big(\!\varrho_{1,i}\mathcal{D}^{2}\!\left(\|((\pmb{l}_{j^{U}}^{*})^{(n+1)}\!,\!h_{j^{U}})\!\!-\!\!\pmb{s}_{i}\|_{2}^{2}\right) \!\!\\
&\! +\!\varrho_{2,i,k}\mathcal{D}\!\left(\|((\pmb{l}_{j^{U}}^{*})^{(n+1)},h_{j^{U}})\!\!-\!\!\pmb{s}_{i}\|_{2}^{2}\right)\!\!+\!\!\varrho_{3,i,k}\Big),~~b_{i,j^{U}}\!\!=\!\!1,
\end{aligned}
\right.
\end{equation}
where $\mathcal{D}(x)\!\triangleq\!\kappa_{0}^{2}x\left[\frac{\mu^{\rm{LoS}}\!-\!\mu^{\rm{NLoS}}}{1\!+\!\beta \exp\left(-\psi\left[\frac{180}{\pi}\arcsin\left(\frac{h_{j^{U}}}{\sqrt{x}}\right)\!-\!\beta\right]\right)}\!+\!\mu^{\rm{NLoS}}\right]$.

The suboptimal altitude of UAV is then obtained from the argument that minimizes the following one-dimensional function as follows:
\begin{equation}\underline{\textbf{P3b-2:}}\label{P3b-2}~~~~
(h_{j^{U}}^*)^{n+1}=\arg\max_{h_{j^{U}}} \sum_{k=1}^{L_{j^{U}}}\sum_{i=1}^{C_{j^{U}}^{k}}f_{i,j^{U}}^{3b-2},
\end{equation}
where $\sum_{k=1}^{L_{j^{U}}}\sum_{i=1}^{C_{j^{U}}^{k}}f_{i,j^{U}}^{3b-2}$ is a quasi-concave function when both $b_{i,j^{U}}=0$ and $b_{i,j^{U}}=1$ as verified by numerical simulation based on the parameters in Table \ref{tabel_SS}. The suboptimal altitude of the UAV is then obtained via one-dimensional search over a feasible range of altitudes.

After $(\pmb{u}_{j^{U}}^{*})^{(n+1)}=((x_{j^{U}}^{*})^{(n+1)},(y_{j^{U}}^{*})^{(n+1)},(h_{j^{U}}^{*})^{(n+1)})$ is obtained in the $(n\!+\!1)$th iteration, an efficient gradient descent algorithm can be applied to obtain the optimal value of $\xi_{i}^{(n+1)}$ by letting the gradient of objective function \eqref{1-3b_determine} respect to $\xi_{i}$ equals to a zero. The resultant $\xi_{i}^{(n+1)}$ is as follows:
\begin{equation}\label{xi_update}
\xi_{i}^{(n+1)}\!\!=\!\!\left\{
\begin{aligned}
&\frac{\sqrt{\varphi_{1,i,k}}}{\varrho_{1,i}\bar{D}_{i,j^{U}}\left(\resizebox{0.095\hsize}{!}{$(\pmb{u}_{j^{U}}^{*})^{(n+1)}$}\right) +  \varrho_{2,i,k}},~~b_{i,j^{U}}\!\!=\!\!0,\\
&\frac{\sqrt{\varphi_{1,i,k} \bar{D}_{i,j^{U}}\left(\resizebox{0.095\hsize}{!}{$(\pmb{u}_{j^{U}}^{*})^{(n+1)}$}\right)\!\!+\!\!\varphi_{2,i,k}} }{(\varrho_{1,i}\bar{D}_{i,j^{U}}\left(\resizebox{0.095\hsize}{!}{$(\pmb{u}_{j^{U}}^{*})^{(n+1)}$}\right) \!\!+\!\!  \varrho_{2,i,k})\bar{D}_{i,j^{U}}\left(\resizebox{0.095\hsize}{!}{$(\pmb{u}_{j^{U}}^{*})^{(n+1)}$}\right)\!\!+\!\!\varrho_{3,i,k}},~~b_{i,j^{U}}\!\!=\!\!1.
\end{aligned}
\right.
\end{equation}

Similarly, \eqref{P3b-1}, \eqref{P3b-2}, and \eqref{xi_update} are guaranteed to achieve a stationary point of concave-convex fractional programming problems with a nondecreasing sum-of-functions-of-ratio value in each iteration. In the end, the value of $\varsigma_{i}$ can be reset through an explore method, and the globally optimal solution can be obtained after the multiple comparisons.

To solve the original optimization problem \textbf{OP}, the time allocation and scheduling order (presented in subsection III-A), the DL device association and UAVs’ locations (optimization in III-B), and the UL device association and UAVs’ locations (optimization in III-C) are applied iteratively until there is no change in sum throughput of the network. Evidently, at each iteration, the sum throughput of the system increases monotonically. Hence, the suboptimal solution converges a locally optimum after several iterations. The main procedure for solving \textbf{OP} is summarized in Algorithm \ref{algorithm-sum}. In step 15, the sum throughput of the last output of the algorithm is compared with the sum throughput of the previous iteration. When the growth is less than a certain threshold, it can be determined that the optimal value has no longer changed, that is, the convergence is reached.


\begin{algorithm}[htb]
\caption{\small{Main steps for the optimal solutions of \textbf{OP}}} \label{algorithm-sum}
{\footnotesize{
\begin{algorithmic}[1]
\STATE Initialize $n=0$, $\pmb{u}_{j^{D}}^{(0)}$, $\pmb{u}_{j^{U}}^{(0)}$, $\pmb{s}$, $\textbf{I}^{(0)}$, $\textbf{A}^{(0)}$, $\textbf{B}^{(0)}$, $\pmb{\tau}^{(0)}$.
\STATE Repeat:
\STATE $n=n+1$.
\STATE Compute $(\pmb{w}^{*})^{(n)}$ from \eqref{theorem-w_ik} under $(\pmb{u}_{j^{D}}^{*})^{(n\!-\!1)}$, $(\pmb{u}_{j^{U}}^{*})^{(n\!-\!1)}$, $(\textbf{I}^{*})^{(n\!-\!1)}$, $(\textbf{A}^{*})^{(n\!-\!1)}$, $(\textbf{B}^{*})^{(n\!-\!1)}$, $(\pmb{\tau}^{*})^{(n\!-\!1)}$.
\STATE Compute $(\pmb{s}^{*})^{(n)}$ from \eqref{s_ik} and \eqref{theorem-A_juk} under $(\pmb{u}_{j^{D}}^{*})^{(n\!-\!1)}$, $(\pmb{u}_{j^{U}}^{*})^{(n\!-\!1)}$, $(\textbf{I}^{*})^{(n\!-\!1)}$, $(\textbf{A}^{*})^{(n\!-\!1)}$, $(\textbf{B}^{*})^{(n\!-\!1)}$, $(\pmb{\tau}^{*})^{(n\!-\!1)}$, $(\pmb{w}^{*})^{(n)}$.
\STATE Compute $(\pmb{\tau}^{*})^{(n)}$ via one-dimensional search over $(0,T_{\rm{hov}})$.
\STATE Compute $\varpi\!\!=\!\!\max\{\frac{\tau_{1}^{*}s_{i,k}^{*}\gamma}{\Theta_{i}^{0}L_{j^{U}}\tau_{0}^{*}\!+\!\Theta_{i}^{1}(k\!-\!1)\tau_{1}^{*}}\}$ under $(\pmb{u}_{j^{D}}^{*})^{(n\!-\!1)}$, $(\pmb{u}_{j^{U}}^{*})^{(n\!-\!1)}$, $(\textbf{I}^{*})^{(n\!-\!1)}$, $(\textbf{A}^{*})^{(n\!-\!1)}$, $(\textbf{B}^{*})^{(n\!-\!1)}$, $(\pmb{s}^{*})^{(n)}$, $(\pmb{\tau}^{*})^{(n)}$.
\IF{$ \varpi<1 $}
\STATE Update $(\textbf{I}^*)^{(n)}$ and $(\pmb{u}_{j^{D}}^*)^{(n)}$ from Section \ref{Sec-DL} by bringing in $\tau_{1}^{*}$, $\tau_{0}^{*}$ and $(\textbf{A}^*)^{(n\!-\!1)}$, $(\textbf{B}^*)^{(n\!-\!1)}$ and $(\pmb{u}_{j^{U}}^*)^{(n\!-\!1)}$.
\STATE Update $(\textbf{A}^*)^{(n)}$, $(\textbf{B}^*)^{(n)}$ and $(\pmb{u}_{j^{U}}^*)^{(n)}$ from Section \ref{Sec_UL} by bringing in $\tau_{1}^{*}$, $\tau_{0}^{*}$ and $(\textbf{I}^*)^{(n)}$ and $(\pmb{u}_{j^{D}}^*)^{(n)}$.
\ELSE
\STATE Update $(\textbf{I}^*)^{(n)}$ and $(\pmb{u}_{j^{D}}^*)^{(n)}$ from Section \ref{Sec-DL} by bringing in $\Theta_{m}$ and $(\textbf{A}^*)^{(n\!-\!1)}$, $(\textbf{B}^*)^{(n\!-\!1)}$ and $(\pmb{u}_{j^{U}}^*)^{(n\!-\!1)}$.
\STATE Update $(\textbf{A}^*)^{(n)}$, $(\textbf{B}^*)^{(n)}$ and $(\pmb{u}_{j^{U}}^*)^{(n)}$ from Section \ref{Sec_UL} by bringing in $\Theta_{m}$ and $(\textbf{I}^*)^{(n)}$ and $(\pmb{u}_{j^{D}}^*)^{(n)}$.
\ENDIF
\STATE Until: Convergence of sum throughput.
\STATE Output: $\pmb{\tau}^{*}$, $\pmb{s}^{*}$, $\textbf{A}^*$, $\textbf{B}^*$, $\textbf{I}^*$ and $\pmb{u}_{j^{D}}^*$,$\pmb{u}_{j^{U}}^*$.
\end{algorithmic}}}
\end{algorithm}

For UL and DL optimization, the complexity of the worst case is $\mathcal{O}\left(L_{1} N  K^{3}\right)$, where $L_{1}$ denotes the average number of iterations for the convergence of Dinkelbach based algorithm. Thus, the complexity of Algorithm \ref{algorithm-sum} is $\mathcal{O}\left(L_{0} L_{1} N K^{3}\right)$, where $L_{0}$ denotes the average number of iterations for the convergence of Algorithm \ref{algorithm-sum}. For implementation, we consider a centralized network, in which the locations of the devices and UAVs are known to a control center located at a central cloud server. The cloud server will determine the DL and UL time allocation, the UAVs’ locations, the device-UAV association, and the scheduling order of each IoT device.

\section{Numerical Results}\label{Numerical Results}
In this section, we present several numerical examples to evaluate the performance of the proposed algorithm.
\subsection{Simulation Setup}

We consider 80 IoT devices, i.e., $K=80$, which are uniformly located within a circular area with a radius of 80 meters (m), and four UAVs, i.e., $N=4$, in the coverage area. An urban environment is considered with $\beta\!\!=\!\!11.95$ and $\psi\!\!=\!\!0.14$ at 2GHz carrier frequency \cite{6863654}. For NB-IoT, the number of available subcarriers is $M=12$. The results given in the following simulation are averaged over a large number of independent runs. Table \ref{tabel_SS} lists other simulation parameters. In this study, we follow the simulation parameters in the existing studies, which are prevalently used, e.g., \cite{8489918}, \cite{8053918}, and \cite{8038869}.

\begin{table}[t]
\setlength{\abovecaptionskip}{-0.1cm}   
\setlength{\belowcaptionskip}{-2cm}   
 \centering
 \caption{Simulation Parameters}\label{tabel_SS}
\begin{tabular}{|c|c|c|}
\hline
\textbf{Description}&\textbf{Parameter}&\textbf{Value}\\
\hline
EH efficiency under interference&$\eta_i\delta_{i}$& 0.5\\
\hline
Additional path loss for LoS link&$\mu^{\rm{LoS}}$&3~\rm{dB}\\
\hline
Additional path loss for NLoS link&$\mu^{\rm{NLoS}}$&23~\rm{dB}\\
\hline
Noise power &$N_{0}$&-120~\rm{dBm}\\
\hline
DL EH threshold&$\rho$&-18~\rm{dBm} in \cite{6613706}\\
\hline
UL SNR threshold&$\gamma$&5~\rm{dB}\\
\hline
\end{tabular}
\vspace{-0.5cm}
\end{table}

\subsection{EH threshold effect on the location of UAVs}
\begin{figure}[t]
\centering
\subfigure[$\rho=-18~\rm{dBm}$.]{
\label{3D_WET_-18}
\includegraphics[height=6cm]{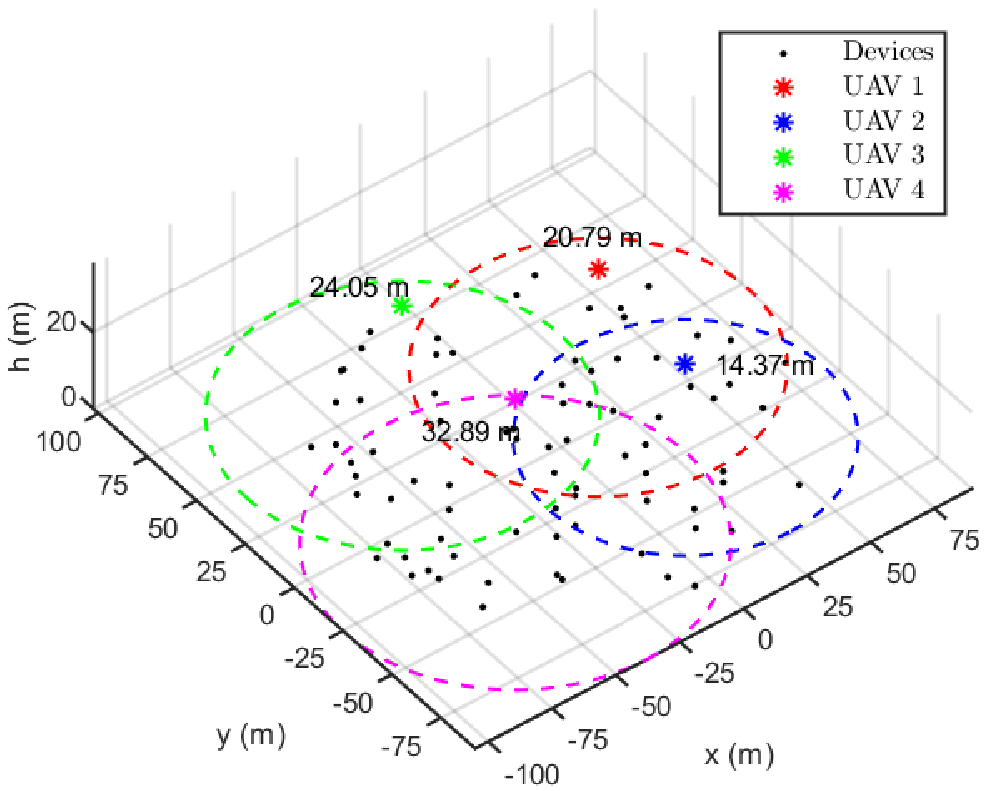}
}
\vspace{-0.15cm}
\quad
\subfigure[$\rho=-28~\rm{dBm}$.]{
\label{3D_WET_-28}
\includegraphics[height=6cm]{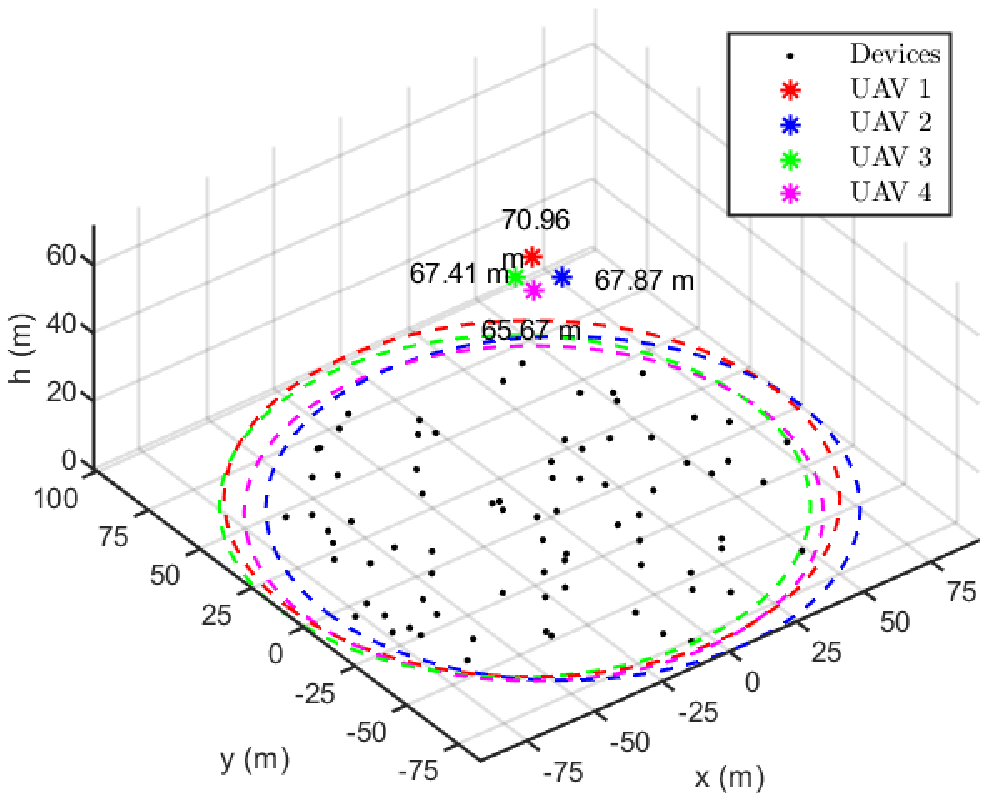}
}
\vspace{-0.15cm}
\quad
\caption{Example of DL energy transfer: the location and coverage of UAVs when $P_{ut}=10~\rm{dB}$.}  \label{3D_WET}
\end{figure}

In Fig. \ref{3D_WET}, we investigate the effect of the EH threshold $\rho$ on the 3D locations of UAVs in DL. From Fig. \ref{3D_WET_-18}, we observe that each UAV adjusts its height and location to cover a part of devices when the EH threshold of the device is relatively high. On the other hand, when the EH threshold is low in Fig. \ref{3D_WET_-28}, each UAV would increase the hovering height and concentrate on the center of the area, i.e., a convergence phenomenon, so that all devices simultaneous harvest energy from multiple UAVs.



\subsection{Effect of the UAVs’ transmit power}

\color{blue}
\begin{figure}[t]
\centering
\subfigure[Altitude and coverage vs. UAVs' transmit power ($P_{ut}$).]{
\label{h_r}
\includegraphics[height=6cm]{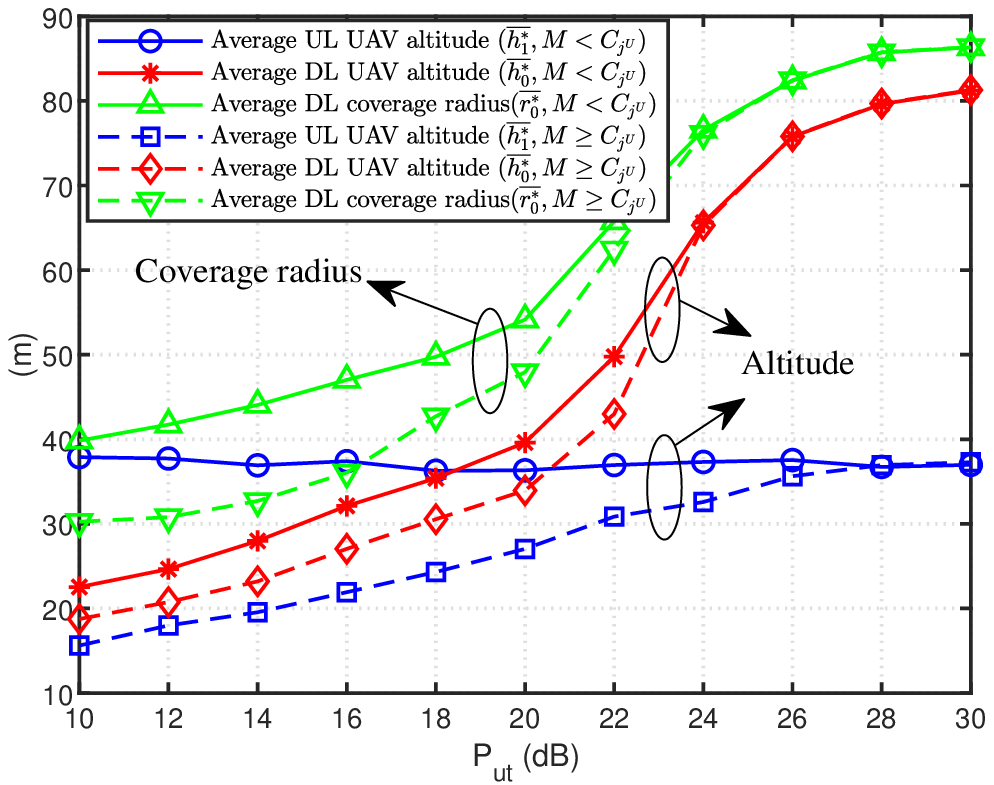}
}
\vspace{-0.15cm}
\quad
\subfigure[Throughput vs. UAVs' transmit power ($P_{ut}$).]{
\label{Sum_throughput}
\includegraphics[height=6cm]{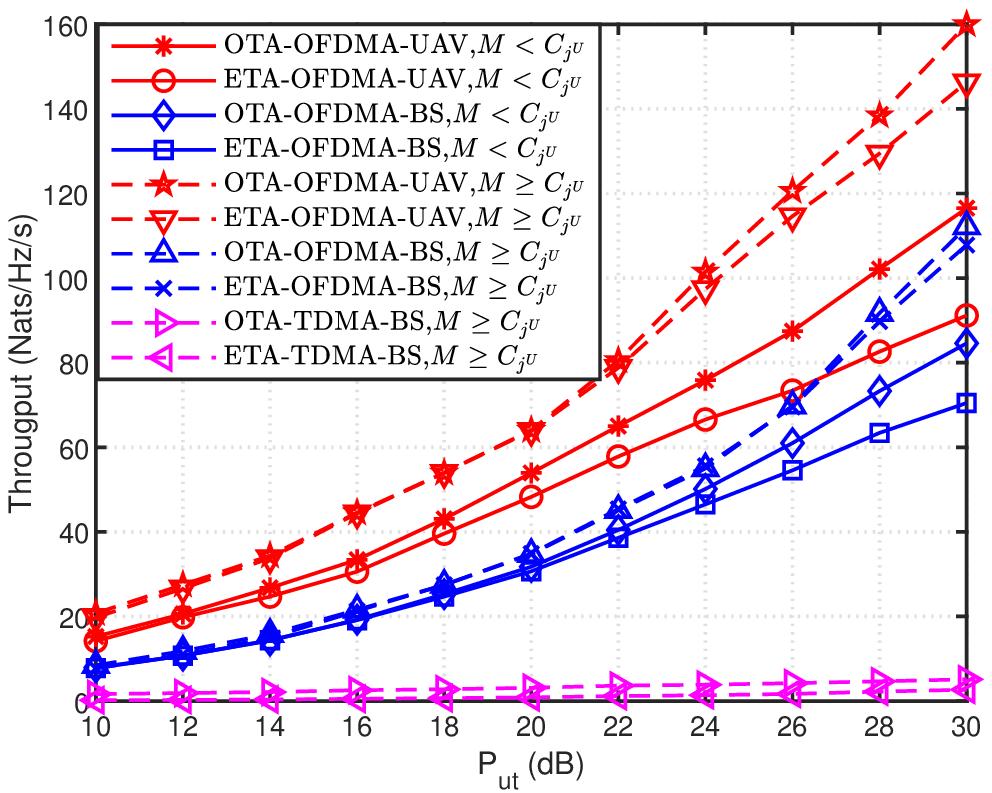}
}
\vspace{-0.15cm}
\quad
\caption{Effect of the UAVs’ transmit power.} \label{OTA-P}
\vspace{-0.5cm}
\end{figure}\color{black}
There is a special case. As the number of channels increases, the number of epochs gradually decreases and tends to one. When $M \geq C_{j^{U}}$, the full-duplex mode of the network degrades to the half-duplex mode of the network in \cite{9013224}, and all devices could simultaneously send their information to the assigned UAVs in UL by using harvested energy from UAVs in DL. To evaluate the performance differences between two modes, all of the subsequent simulations will be performed under two conditions, namely, $M < C_{j^{U}}$ and $M \geq C_{j^{U}}$.

In Fig. \ref{h_r}, we investigate the effect of the transmit power on the altitudes of UAVs in UL and DL and coverage range of UAVs in DL. It is observed that the average altitude and coverage radius of UAVs increase up to a certain level and are saturated as the transmit power of UAVs increases. The average converge radius of UAVs gradually converges to about $90~\rm{m}$, because the ground devices are uniformly distributed within a fixed area with a radius of $80~\rm{m}$. Also, when the transmit power is lower than 22 \rm{dB}, the average height and coverage radius of DL when the number of subcarriers $M \geq C_{j^{U}}$ are lower than those of $M < C_{j^{U}}$. Because all devices only harvest energy in the DL when $M \geq C_{j^{U}}$, the average height and coverage radius of UAVs are reduced so that the devices can harvest enough energy to satisfy the UL SNR requirement. However, when the transmission power is higher than 22 \rm{dB}, the average altitude and coverage radius of DL when $M \geq C_{j^{U}}$ and those of $M < C_{j^{U}}$ are almost equal. Because each device can harvest energy from multiple UAVs when $P_{ut}$ is sufficiently large, the UAVs under conditions that $M \geq C_{j^{U}}$ and $M < C_{j^{U}}$ both are coincided in the center of the converage owing to the convergence phenomenon. Furthermore, it is also observed that the UL average altitude is almost stable at approximately 40 m when $M < C_{j^{U}}$. Because the UAV performs the full-duplex mode and UAV still needs to improve its coverage radius for devices scheduled in the later epochs to harvest more energy. However, when $M \geq C_{j^{U}}$, the average altitude of UL gradually increases from $18~\rm{m}$ to $40~\rm{m}$. This means that the transmit power has more impact on the UL average altitude of UAVs especially when the number of channels is large.


In Fig. \ref{Sum_throughput}, the proposed scheme, denoted by OFDMA-UAV is compared to the fixed BS schemes, in terms of the network throughput. Here, the locations of BSs are computed by so-called a disk covering problem, and each BS covers area radius is $56~\rm{m}$$(\simeq80~\rm{m}\times\sqrt{2}/2)$ according to four disks formula in \cite{kershner1939number}. The heights of BSs are set by $H\!\!=\!\!40~\rm{m}$ to achieve a coverage radius of $56~\rm{m}$ while adopting the optimal DL and UL device association of Subsections III-B and III-C. The optimal time allocation (OTA) and equal time allocation (ETA) strategies are investigated for the comparison. For the sake of comparison when $M \geq C_{j^{U}}$, the existing TDMA-BS scheme in \cite{6678102} is also compared. From the results, it is verified that the proposed OTA-OFDMA-UAV ($M \geq C_{j^{U}}$) achieves the highest throughput. Moreover, it is observed that the OFDMA-based schemes significantly outperform the TDMA-based schemes when $M \geq C_{j^{U}}$. From an observation that the throughput gap between the UAVs scheme and the BSs scheme increases as $P_{ut}$ increases, we can surmise that location optimization with high $P_{ut}$ has more impact on throughput improvement. It is evident to observe that the performances of OTA-OFDMA-UAV (BS) and ETA-OFDMA-UAV (BS) in $M < C_{j^{U}}$ are worse than the performances of those in $M \geq C_{j^{U}}$ due to the limited number of channels. Also, it is observed that OTA slightly outperforms ETA, and the throughput gap decreases and turns to increase as $P_{ut}$ increases when $M \geq C_{j^{U}}$. Since OTA outperforms slightly ETA when $M \geq C_{j^{U}}$, we conclude that the ETA strategy can be a suitable substitute for OTA to reduce the optimization complexity when the number of channels is sufficiently large. Furthermore, we can observe that the gap between the OTA and ETA schemes increases as $P_{ut}$ increases when $M < C_{j^{U}}$. This indicates that time allocation is more important when $P_{ut}$ is large and number of channels is limited.

\subsection{Effect of the User Scheduling}
\begin{figure}[t]
\centering
\subfigure[Throughput vs. UAVs' transmit power ($P_{ut}$).]{
\label{T_P3}
\includegraphics[height=6cm]{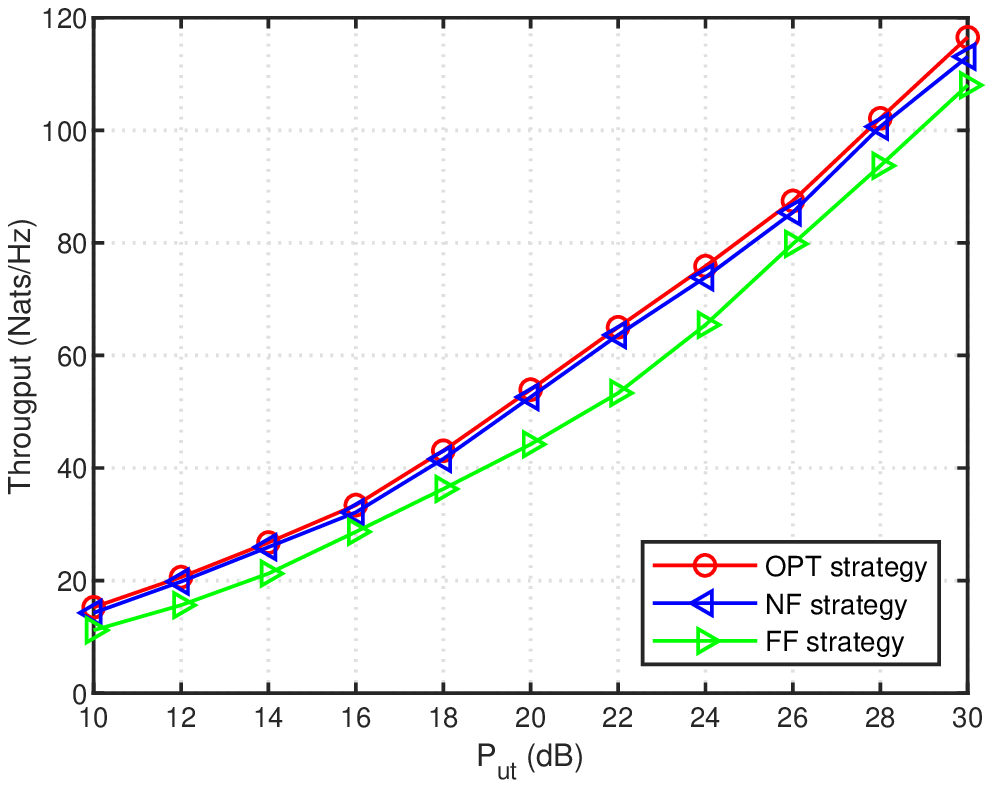}
}
\vspace{-0.15cm}
\quad
\subfigure[System fairness vs. UAVs' transmit power ($P_{ut}$).]{
\label{J_P3}
\includegraphics[height=6cm]{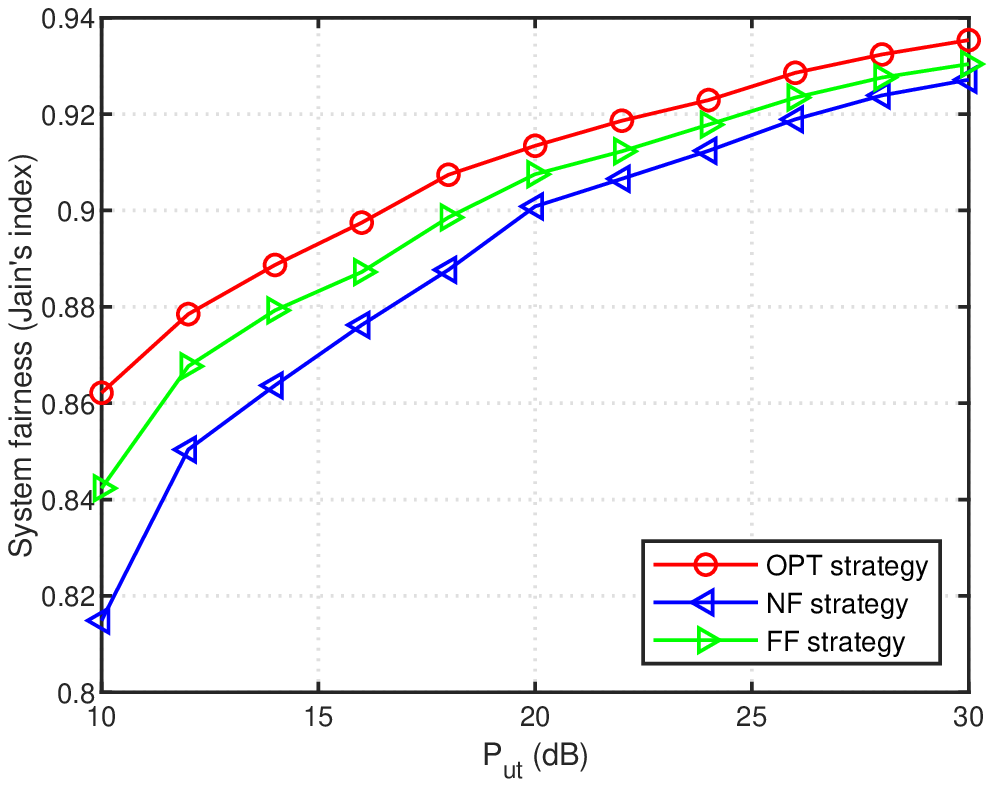}
}
\vspace{-0.15cm}
\quad
\caption{Effect of the User Scheduling.} \label{OTA-S}
\vspace{-0.5cm}
\end{figure}

Here, we propose two suboptimal device scheduling schemes for comparison, which are given as follows.

Firstly, a circle is drawn with the horizontal position of UAV as the center, and this circle covers $C_{j^{U}}$ devices served by UAV in UL. Similarly, a smaller circle is drawn with the same center, such that the number of devices covered by the ring donated by $\mathcal{A}^{1}$ between the two circles is $M$. Keep performing this procedure by moving toward inside and denoting the rings by $\mathcal{A}^{2},...,\mathcal{A}^{L_j^U-1}$, sequentially, until the number of devices covered by the innermost circle donated by $\mathcal{A}^{L_j^U}$ is less than or equal to $M$.

(i) Far-First (FF) Strategy:  In the UL phase, $M$ devices located in the outermost ring, i.e., $\mathcal{A}^{1}$,  first perform UL communication at epoch 1. $M$ devices covered by the second ring from outside, i.e., $\mathcal{A}^{2}$,  then perform UL communication at epoch 2. By moving toward inside and repeating the same procedure until $(C_{j^{U}}\!-\!M(L_{j^{U}}\!-\!1))$ devices located in the innermost circle, i.e., $\mathcal{A}^{L_j^U}$, complete the UL communication at the last epoch $L_{j^{U}}$.

(ii) Near-First (NF) Strategy: In the UL phase, $(C_{j^{U}}\!-\!M(L_{j^{U}}\!-\!1))$ devices located in the innermost circle, i.e., $\mathcal{A}^{L_j^U}$, first performs UL communication at epoch 1, followed by $M$ devices covered by the first inner ring, i.e., $\mathcal{A}^{L_j^U-1}$, transmit information at epoch 2. This procedure is repeated until $M$ devices located in the outermost ring, i.e., $\mathcal{A}^{1}$ complete the UL communication at the last epoch $L_{j^{U}}$.

Fig. \ref{T_P3} shows the sum throughput versus transmit power $P_{ut}$ under different device scheduling strategies. It is observed that the proposed optimal scheduling strategy outperforms the NF strategy and FF strategy. Here, note that, for the NF strategy, the number of devices scheduled in the last epoch is $M$. On the other hand, for the FF strategy, the number of devices scheduled in the last epoch is $(C_{j^{U}}-M(L_{j^{U}}-1))$, which is less than or equal to $M$. Therefore, the NF strategy is better than the FF one in terms of throughput. Also, we can observe that the gap between the optimal strategy and the NF strategy is much smaller than that between the optimal strategy and FF strategy. Thus, in practice, we can adapt the NF strategy, which could considerably reduce scheduling overhead while delivering near-optimal overall performance.

Besides, we want to quantify the system fairness for three scheduling schemes by using Jain's fairness index. The Jain's index is defined as $J\!\!=\!\!\left(\!\sum_{i=1}^{K}\mathcal{T}_{i,j^{U}}\!\right)^{2}\!\!\!/\!\!\left(\!K\sum_{i=1}^{K}\mathcal{T}_{i,j^{U}}^{2}\!\right)$, which is bounded in $[1/K, 1]$ \cite{jain1999throughput}.

Fig. \ref{J_P3} shows Jain's index $J$ versus transmit power $P_{ut}$ under different device scheduling strategies. Jain's index increases monotonically with the transmit power $P_{ut}$. In other words, as $P_{ut}$ increases, more fair time allocation, device-and-UAV association, and locations of UAVs in UL and DL can be obtained. The reason for this phenomenon is that the influence of different path loss for different devices can be neglected when transmit power increases. It is also observed that the proposed optimal strategy always has the highest Jain's index, followed by the FF strategy. The NF strategy achieves the lowest Jain's index, because only UL UAV location information is used for scheduling in the NF and FF strategies. Note that our proposed scheduling optimal strategy utilizes both UL and DL UAVs locations and device-and-UAV association information.


\section{Conclusion} \label{Sec-Conclusions}
In this paper, we have proposed a novel framework for FD-OFDMA based UAV-enabled wireless-powered IoT networks, where a swarm of UAVs is deployed in 3D to simultaneously charge all devices and then fly to new locations to collect information from scheduled devices during several epochs via OFDMA. Under the proposed model, we jointly optimized the UL-and-DL 3D deployment of the UAV swarm, including the device-UAV association, the IoT device scheduling order, and the UL-DL time allocation, to maximize the UL sum throughput. The results show that the 3D position of the UAVs have different trends during UL and DL. We also proposed two suboptimal scheduling strategies, referred to as the near-first (NF) scheme and far-first (FF) scheme, by exploiting the system characteristics. It is shown that the proposed suboptimal schemes can achieve a satisfactory performance. It is also observed that the NF scheme outperforms the FF scheme in terms of throughput maximization, but the FF scheme outperforms the NF scheme in terms of fairness. Also, the simulations results show that the number of channels for NB-IoT has a significant influence on UAVs' altitude. When the number of channels is less than the number of devices, the altitude of full-duplex UAVs during UL communications is fixed. On the other hand, when the number of channels is larger than or equal to the number of devices, the UAV operates in a half-duplex mode, and the UL and DL altitudes are lower than that in full-duplex mode.

\appendices
\renewcommand{\theequation}{\thesection.\arabic{equation}}
\setcounter{equation}{0}
\section{Proof of Theorm \ref{theorm-1-time}} \label{proof-1-time}
Although $s_{i,k}$ is a binary variable, we relax the constraint \eqref{1_constraint_tau} as $0\leq s_{i,k}\leq 1$ by time-sharing condition \cite{1658226}. As a result, $s_{i,k}$ can be interpreted as a time-sharing factor in allocating device $i$ to $L_{j^{U}}$ epochs for transmitting information. Actually, for any fixed set of $\tau_{q}$ and $s_{i,k}$, the objective function of \eqref{FW_subP1} is jointly concave, and all the constraints are affine in $\tau_{q}$ and $s_{i,k}$. Thus, according to \cite{boyd2004convex}, we conclude that  \eqref{FW_subP1} has zero duality gap.
The Lagrangian of \textbf{P1} is
\begin{align}\label{Lang}
&\mathcal{L}(\boldsymbol{\tau}, \lambda, \pmb{w}, \pmb{u}, \pmb{v}) = \sum_{j^{U}=1}^{N}\sum_{i=1}^{C_{j^{U}}}\sum_{k=1}^{L_{j^{U}}}s_{i,k}\frac{\tau_{1}}{L_{j^{U}}}\mathrm{ln} \left(1 \!+ \! \frac{\Theta_{i}^{0}L_{j^{U}}\tau_{0}\!+\!\Theta_{i}^{1}(k\!-\!1)\tau_{1}}{\tau_{1}} \right) \!-\! \lambda \Big(\tau_{0}+\tau_{1} \!-\! 1 \Big)\\
&\!-\!\sum_{j^{U}=1}^{N}\sum_{i=1}^{C_{j^{U}}}\sum_{k=1}^{L_{j^{U}}}  w_{i,k}\Big(\frac{\tau_{1}s_{i,k}\gamma}{\Theta_{i}^{0}L_{j^{U}}\tau_{0}\!+\!\Theta_{i}^{1}(k\!-\!1)\tau_{1}}\!-\!1\Big)\!-\!\sum_{i=1}^{C_{j^{U}}}u_{i}(\sum_{k=1}^{L_{j^{U}}}s_{i,k}\!-\!1)\!-\!\sum_{k=1}^{L_{j^{U}}}v_{k}(\sum_{i=1}^{C_{j^{U}}}s_{i,k}\!-\!M),\notag
\end{align}
where $\lambda$, $\pmb{w}$, $\pmb{u}$ and $\pmb{v}$ are the non-negative Lagrange multiplier.

Similar in \cite{8424210,4786456}, \eqref{Lang} can be solved by using its Karush-Kuhn-Tucker (KKT) conditions as follows:
\begin{align}
&\lambda^* \Big(\tau_{0}^*+\tau_{1}^* - 1 \Big) = 0,\label{KKT_sum_T}\\
&w_{i,k}^{*}\Big(\frac{\tau_{1}^{*}s_{i,k}^{*}\gamma}{\Theta_{i}^{0}L_{j^{U}}\tau_{0}^{*}+\Theta_{i}^{1}(k-1)\tau_{1}^{*}}-1\Big) = 0,~\forall i \in \mathcal{K},\label{KKT_SNR}\\
&v_{k}^{*}\Big(\sum_{i=1}^{C_{j^{U}}}s_{i,k}^{*}-M\Big)=0,\label{KKT_I1}\\
&\frac{\partial \mathcal{L}(\tau_{0},\tau_{1}, \lambda^*, \pmb{w}^*, \pmb{u}^*, \pmb{v}^*)}{\partial s_{i,k}} \Big|_{s_{i,k}=s_{i,k}^{*} } = 0,~\forall i \in \mathcal{K},~\forall k \in \mathcal{L}_{j^{U}},\label{1-L-Derivative-I}\\
&\frac{\partial \mathcal{L}(\tau_{0},\tau_{1}, \lambda^*, \pmb{w}^*, \pmb{u}^*, \pmb{v}^*)}{\partial \tau_q} \Big|_{\tau_q=\tau_q^* } = 0,~\forall q \in  \{0,1\},\label{1-L-Derivative-tau}
\end{align}
where $\tau_{0}^*$ and $\tau_{1}^*$ denote the optimal time solution of \textbf{P1}.
Here, \eqref{1-L-Derivative-I} and \eqref{1-L-Derivative-tau} can be further written as follows:
\begin{align}
& \sum_{j^{U}=1}^{N}\sum_{i=1}^{C_{j^{U}}} \frac{\tau_{1}^{*}}{L_{j^{U}}}\mathrm{ln} \left(1 + \frac{\Theta_{i}^{0}L_{j^{U}}\tau_{0}^{*}+\Theta_{i}^{1}(k-1)\tau_{1}^{*}}{\tau_{1}^{*}} \right)-\frac{w_{i,k}^{*}\tau_{1}^{*}\gamma}{\Theta_{i}^{0}L_{j^{U}}\tau_{0}^{*}+\Theta_{i}^{1}(k-1)\tau_{1}^{*}}=u_{i}^{*}+v_{k}^{*},\label{1-L_{j^{U}}-Derivative-I}\\
& \sum_{j^{U}=1}^{N}\sum_{i=1}^{C_{j^{U}}} \sum_{k=1}^{L_{j^{U}}}\frac{s_{i,k}^{*}\Theta_{i}^{0}L_{j^{U}}\tau_{1}^{*}}{\Theta_{i}^{0}L_{j^{U}}\tau_{0}^{*}+[1+\Theta_{i}^{1}(k-1)]\tau_{1}^{*}}  = \lambda^*-\sum_{j^{U}=1}^{N}\sum_{i=1}^{C_{j^{U}}}\sum_{k=1}^{L_{j^{U}}}\frac{w_{i,k}^{*}\tau_{1}^{*}s_{i,k}^{*}\gamma\Theta_{i}^{0}L_{j^{U}}}{[\Theta_{i}^{0}L_{j^{U}}\tau_{0}^{*}+\Theta_{i}^{1}(k-1)\tau_{1}^{*}]^{2}},\label{1-L_{j^{U}}-Derivative-0}\\
& \sum_{j^{U}=1}^{N}\sum_{i=1}^{C_{j^{U}}} \sum_{k=1}^{L_{j^{U}}}s_{i,k}^{*}\left(\mathrm{ln}\left(1 + \frac{\Theta_{i}^{0}L_{j^{U}}\tau_{0}^{*}+\Theta_{i}^{1}(k-1)\tau_{1}^{*}}{\tau_{1}^{*}}\right)-\frac{\Theta_{i}^{0}L_{j^{U}}\tau_{0}^{*}}{\Theta_{i}^{0}L_{j^{U}}\tau_{0}^{*}+[1+\Theta_{i}^{1}(k-1)]\tau_{1}^{*}}\right)\notag\\
& = \lambda^*+\sum_{j^{U}=1}^{N}\sum_{i=1}^{C_{j^{U}}}\sum_{k=1}^{L_{j^{U}}}\frac{w_{i,k}^{*}\tau_{0}^{*}s_{i,k}^{*}\gamma\Theta_{i}^{0}L_{j^{U}}}{[\Theta_{i}^{0}L_{j^{U}}\tau_{0}^{*}+\Theta_{i}^{1}(k-1)\tau_{1}^{*}]^{2}}. \label{1-L_{j^{U}}-Derivative-1}
\end{align}
\subsection{Part 1}
Firstly, we focus only on \eqref{1-L_{j^{U}}-Derivative-I} and show that optimal solution yields a scheduling policy with multi IoT devices selection by adopting a similar approach as in \cite{6661329}. Device $i$ is assigned into epoch $k$ when the following selection criterion is satisfied:
\begin{equation}\label{proof_A_juk}
\mathcal{A}_{j^{U}}^{k} \!\!=\!\!\arg\!\!\!\!\!\!\!\!\!\!\max_{\substack{\mathcal{C}_{j^{U}}^{(S)}\subset\left(\mathcal{C}_{j^{U}}/\sum_{n=k+1}^{L_{j^{U}}}{A}_{j^{U}}^{n}\right), \\
|\mathcal{C}_{j^{U}}^{(S)}|\leq M}}\!\!\sum_{i\in \mathcal{C}_{j^{U}}^{(S)}}\!\!\frac{\tau_{1}^{*}}{L_{j^{U}}}\mathrm{ln}\! \left(\!1 \!+\! \frac{\Theta_{i}^{0}L_{j^{U}}\tau_{0}^{*}\!+\!\Theta_{i}^{1}(k\!-\!1)\tau_{1}^{*}}{\tau_{1}^{*}} \!\right)\!-\!\frac{w_{i,k}^{*}\tau_{1}^{*}\gamma}{\Theta_{i}^{0}L_{j^{U}}\tau_{0}^{*}+\Theta_{i}^{1}(k\!-\!1)\tau_{1}^{*}},
\end{equation}
where $\mathcal{A}_{j^{U}}^{k}$ is the device set that transmits information to UAV $j^{U}$ at epoch $k$. When device $i$ belong to the subset $\mathcal{A}_{j^{U}}^{k}$, $\frac{\tau_{1}^{*}}{L_{j^{U}}}\mathrm{ln} \left(1 + \frac{\Theta_{i}^{0}L_{j^{U}}\tau_{0}^{*}+\Theta_{i}^{1}(k-1)\tau_{1}^{*}}{\tau_{1}^{*}} \right)-\frac{w_{i,k}^{*}\tau_{1}^{*}\gamma}{\Theta_{i}^{0}L_{j^{U}}\tau_{0}^{*}+\Theta_{i}^{1}(k-1)\tau_{1}^{*}}$ is the marginal benefit provided to the system. In other words, device $i\in\mathcal{A}_{j^{U}}^{k}$ is selected for information transmission at epoch $k$ if it can provide the maximum marginal benefit to the system. Besides, $\frac{w_{i,k}^{*}\tau_{1}^{*}\gamma}{\Theta_{i}^{0}L_{j^{U}}\tau_{0}^{*}+\Theta_{i}^{1}(k-1)\tau_{1}^{*}}$ is a penalty function of \eqref{proof_A_juk}, if device $i$ has harvested enough energy to satisfy the UL SNR constraint requirement on \eqref{11_constraint_SNR}, then $w_{i,k}^{*}$ will be equal to 0 due to the complementary slackness condition and the network dispatch center will have a higher preference to make device $i$ served by UAV $j^{U}$ at epoch $k$. Thus, we can set $w_{i,k}^{*}$ as follows:
\begin{equation}\label{w_ik}
w_{i,k}^{*}=\left\{
\begin{aligned}
&1,~\frac{\tau_{1}^{*}\gamma}{\Theta_{i}^{0}L_{j^{U}}\tau_{0}^{*}+\Theta_{i}^{1}(k\!-\!1)\tau_{1}^{*}}\geq1,\\
&0,~\frac{\tau_{1}^{*}\gamma}{\Theta_{i}^{0}L_{j^{U}}\tau_{0}^{*}+\Theta_{i}^{1}(k\!-\!1)\tau_{1}^{*}}<1.
\end{aligned}
\right.
\end{equation}

\subsection{Part 2}
\textit{\textbf{Case 1:}} $\frac{\tau_{1}^{*}s_{m,n}^{*}\gamma}{\Theta_{m}^{0}L_{j^{U}}\tau_{0}^{*}+\Theta_{m}^{1}(n-1)\tau_{1}^{*}}<1$ where $[m,n]=\arg\max_{[i \in  \mathcal{K}, k\in \mathcal{L}_{j^{U}}]}\frac{\tau_{1}^{*}s_{i,k}^{*}\gamma}{\Theta_{i}^{0}L_{j^{U}}\tau_{0}^{*}+\Theta_{i}^{1}(k-1)\tau_{1}^{*}}$.

In this case, $s_{i,k}^{*}w_{i,k}^{*}=0$, $~\forall i \in  \mathcal{C}_{j^{U}}$, $\forall k \in  \mathcal{L}_{j^{U}}$, and $\forall j^{U} \in  \mathcal{N}$, from the complementary slackness conditions, \eqref{KKT_SNR}. The term $\sum_{j^{U}=1}^{N}\sum_{i=1}^{C_{j^{U}}}\sum_{k=1}^{L_{j^{U}}}\frac{w_{i,k}^{*}\tau_{0}^{*}s_{i,k}^{*}\gamma\Theta_{i}^{0}L_{j^{U}}}{[\Theta_{i}^{0}L_{j^{U}}\tau_{0}^{*}+\Theta_{i}^{1}(k-1)\tau_{1}^{*}]^{2}}$ in \eqref{1-L_{j^{U}}-Derivative-0} and \eqref{1-L_{j^{U}}-Derivative-1} can then be omitted; as a result, it is observed that the right hand sides of \eqref{1-L_{j^{U}}-Derivative-0} and \eqref{1-L_{j^{U}}-Derivative-1} are identical to each other. Thus, by substituting \eqref{1-L_{j^{U}}-Derivative-0} and \eqref{1-L_{j^{U}}-Derivative-1}, we have
\begin{equation}\label{1-L_{j^{U}}-Derivative-01}
  \sum_{j^{U}=1}^{N}\sum_{i=1}^{C_{j^{U}}}\sum_{k=1}^{L_{j^{U}}}s_{i,k}^{*} \mathrm{ln}(1 \!+\! \frac{\Theta_{i}^{0}L_{j^{U}}\tau_{0}^{*}\!\!+\!\!\Theta_{i}^{1}(k\!\!-\!\!1)\tau_{1}^{*}}{\tau_{1}^{*}}) \!\!=\!\! \sum_{j^{U}=1}^{N}\sum_{i=1}^{C_{j^{U}}}\sum_{k=1}^{L_{j^{U}}} \frac{s_{i,k}^{*} \Theta_{i}^{0}L_{j^{U}}(\tau_{1}^*\!+\!\tau_{0}^*)}{\Theta_{i}^{0}L_{j^{U}}\tau_{0}^{*}\!+\![1\!+\!\Theta_{i}^{1}(k\!-\!1)]\tau_{1}^{*} }.
\end{equation}

Union \eqref{KKT_sum_T} and \eqref{1-L_{j^{U}}-Derivative-0}, we can know that the optimal time allocation $\tau_0^*$ and $\tau_1^*$ must satisfy $\tau_0^*+\tau_1^*=1$. What's more, it can be easily observed that the left hand side of the \eqref{1-L_{j^{U}}-Derivative-01} can be written as $ \sum_{j^{U}=1}^{N}\sum_{i=1}^{C_{j^{U}}}\sum_{k=1}^{L_{j^{U}}}R_{i,j^{U}}^{k}(\tau_0^*,\tau_1^*)$.

Now \eqref{1-L_{j^{U}}-Derivative-01} can be rewritten as follows:
\begin{equation}
  \sum_{j^{U}=1}^{N}\sum_{i=1}^{C_{j^{U}}}\sum_{k=1}^{L_{j^{U}}}s_{i,k}^{*} R_{i,j^{U}}^{k}(\tau_0^*,\tau_1^*) = \sum_{j^{U}=1}^{N}\sum_{i=1}^{C_{j^{U}}}\sum_{k=1}^{L_{j^{U}}} \frac{s_{i,k}^{*}\Theta_{i}^{0}L_{j^{U}}}{\Theta_{i}^{0}L_{j^{U}}\tau_{0}^{*}+[1+\Theta_{i}^{1}(k-1)]\tau_{1}^{*} }.
\end{equation}

\textit{\textbf{Case 2:}} $\frac{\tau_{1}^{*}s_{m,n}^{*}\gamma}{\Theta_{m}^{0}L_{j^{U}}\tau_{0}^{*}+\Theta_{m}^{1}(n-1)\tau_{1}^{*}}\geq1$

Then we can solve $\tau_{0}^{*}$ and $\tau_{1}^{*}$ by using $\tau_0^*+\tau_1^*=1$ and $\frac{\tau_{1}^{*}\gamma}{\Theta_{m}^{0}L_{j^{U}}\tau_{0}^{*}+\Theta_{m}^{1}(n-1)\tau_{1}^{*}}=1$, which as follows:
\begin{align}
  \tau_{0}^{*}=\frac{\gamma-\Theta_{m}^{1}(n-1)}{\gamma+\Theta_{m}^{0}L_{j^{U}}-\Theta_{m}^{1}(n-1)}, \\
  \tau_{1}^{*}=\frac{\Theta_{m}^{0}L_{j^{U}}}{\gamma+\Theta_{m}^{0}L_{j^{U}}-\Theta_{m}^{1}(n-1)},
\end{align}
which completes the proof.

\bibliographystyle{IEEEtran}
\bibliography{IEEEabrv,ref}

\begin{thebibliography}{10}
\providecommand{\url}[1]{#1}
\csname url@samestyle\endcsname
\providecommand{\newblock}{\relax}
\providecommand{\bibinfo}[2]{#2}
\providecommand{\BIBentrySTDinterwordspacing}{\spaceskip=0pt\relax}
\providecommand{\BIBentryALTinterwordstretchfactor}{4}
\providecommand{\BIBentryALTinterwordspacing}{\spaceskip=\fontdimen2\font plus
\BIBentryALTinterwordstretchfactor\fontdimen3\font minus
  \fontdimen4\font\relax}
\providecommand{\BIBforeignlanguage}[2]{{%
\expandafter\ifx\csname l@#1\endcsname\relax
\typeout{** WARNING: IEEEtran.bst: No hyphenation pattern has been}%
\typeout{** loaded for the language `#1'. Using the pattern for}%
\typeout{** the default language instead.}%
\else
\language=\csname l@#1\endcsname
\fi
#2}}
\providecommand{\BIBdecl}{\relax}
\BIBdecl

\bibitem{9013224}
H.-T. {Ye}, X.~{Kang}, Y.-C. {Liang}, and J.~{Joung}, ``Joint uplink and
  downlink {3D} optimization of an {UAV} swarm for wireless-powered {NB-IoT},''
  in \emph{Proc. Globecom-2019}, Hawaii, USA, Dec. 2019, pp. 1--6.

\bibitem{ALAVI2018589}
\BIBentryALTinterwordspacing
A.~H. Alavi, P.~Jiao, W.~G. Buttlar, and N.~Lajnef, ``Internet of
  things-enabled smart cities: State-of-the-art and future trends,''
  \emph{Measurement}, vol. 129, pp. 589--606, 2018. [Online]. Available:
  \url{http://www.sciencedirect.com/science/article/pii/S0263224118306912}
\BIBentrySTDinterwordspacing

\bibitem{hw}
RWS-180023, ``{3GPP’s Low-Power Wide-Area IoT Solutions:NB-IoT and eMTC},''
  Workshop on 3GPP Submission Towards IMT-2020, Brussels, Belgium, resreport,
  Oct. 2018.

\bibitem{shearer2019powering}
J.~G. Shearer, C.~E. Greene, and D.~W. Harrist, ``Powering devices using {RF}
  energy harvesting,'' May 2019, {US Patent 10,284,019}.

\bibitem{WPT2}
\BIBentryALTinterwordspacing
{Credence Research}, ``Wireless power transmission market forecast,'' Jul.
  2019. [Online]. Available:
  \url{https://www.credenceresearch.com/report/wireless-power-transmission-market.}
\BIBentrySTDinterwordspacing

\bibitem{7984754}
D.~{Niyato}, D.~I. {Kim}, M.~{Maso}, and Z.~{Han}, ``Wireless powered
  communication networks: Research directions and technological approaches,''
  \emph{IEEE Wireless Commun.}, vol.~24, no.~6, pp. 88--97, 2017.

\bibitem{6800126}
X.~Kang, Y.~Chia, C.~K. Ho, and S.~Sun, ``Cost minimization for fading channels
  with energy harvesting and conventional energy,'' \emph{IEEE Trans. Wireless
  Commun.}, vol.~13, no.~8, pp. 4586--4598, Aug. 2014.

\bibitem{6678102}
H.~{Ju} and R.~{Zhang}, ``Throughput maximization in wireless powered
  communication networks,'' \emph{IEEE Trans. Wireless Commun.}, vol.~13,
  no.~1, pp. 418--428, Jan. 2014.

\bibitem{7115936}
X.~Kang, C.~K. Ho, and S.~Sun, ``Full-duplex wireless-powered communication
  network with energy causality,'' \emph{IEEE Trans. Wireless Commun.},
  vol.~14, no.~10, pp. 5539--5551, Oct. 2015.

\bibitem{8489918}
L.~{Xie}, J.~{Xu}, and R.~{Zhang}, ``Throughput maximization for {UAV}-enabled
  wireless powered communication networks,'' \emph{IEEE Internet Things J.},
  vol.~6, no.~2, pp. 1690--1703, Apr. 2019.

\bibitem{8632980}
M.~{Jiang}, Y.~{Li}, Q.~{Zhang}, and J.~{Qin}, ``Joint position and time
  allocation optimization of {UAV} enabled time allocation optimization
  networks,'' \emph{IEEE Trans. Commun.}, vol.~67, no.~5, pp. 3806--3816, May
  2019.

\bibitem{8761608}
H.-T. Ye, X.~Kang, J.~Joung, and Y.-C. Liang, ``Optimal time allocation for
  full-duplex wireless-powered {IoT} networks with unmanned aerial vehicle,''
  in \emph{Proc. ICC-2019}, Shang Hai, May 2019, pp. 1--6.

\bibitem{9080561}
H.-T. {Ye}, X.~{Kang}, J.~{Joung}, and Y.-C. {Liang}, ``Optimization for
  full-duplex rotary-wing {UAV}-enabled wireless-powered {IoT} networks,''
  \emph{IEEE Trans. Wireless Commun.}, vol.~19, no.~7, pp. 5057--5072, 2020.

\bibitem{6863654}
A.~{Al-Hourani}, S.~{Kandeepan}, and S.~{Lardner}, ``Optimal {LAP} altitude for
  maximum coverage,'' \emph{IEEE Wireless Commun. Lett.}, vol.~3, no.~6, pp.
  569--572, Dec. 2014.

\bibitem{3GPP}
{3GPP TR 36.777 V1.0.0}, ``{Enhanced LTE support for aerial vehicles},'' 3GPP,
  Sophia Antipolis, Valbonne, France, Tech. Rep., Dec. 2017.

\bibitem{jain2011practical}
M.~Jain, J.~I. Choi, T.~Kim, D.~Bharadia, S.~Seth, K.~Srinivasan, P.~Levis,
  S.~Katti, and P.~Sinha, ``Practical, real-time, full duplex wireless,'' in
  \emph{Proceedings of the 17th annual international conference on Mobile
  computing and networking}.\hskip 1em plus 0.5em minus 0.4em\relax ACM, 2011,
  pp. 301--312.

\bibitem{8684899}
H.~{Kang}, J.~{Joung}, J.~{Ahn}, and J.~{Kang}, ``Secrecy-aware altitude
  optimization for quasi-static {UAV} base station without eavesdropper
  location information,'' \emph{IEEE Commun. Lett.}, vol.~23, no.~5, pp.
  851--854, May 2019.

\bibitem{8053918}
M.~{Mozaffari}, W.~{Saad}, M.~{Bennis}, and M.~{Debbah}, ``Wireless
  communication using unmanned aerial vehicles ({UAVs}): {Optimal} transport
  theory for hover time optimization,'' \emph{IEEE Trans. Wireless Commun.},
  vol.~16, no.~12, pp. 8052--8066, Dec. 2017.

\bibitem{8038869}
M.~{Mozaffari}, W.~{Saad}, M.~{Bennis}, and M.~{Debbah}, ``Mobile unmanned aerial vehicles ({UAVs}) for energy-efficient
  internet of things communications,'' \emph{IEEE Trans. Wireless Commun.},
  vol.~16, no.~11, pp. 7574--7589, Nov. 2017.

\bibitem{7934322}
K.~{Xiong}, B.~{Wang}, and K.~J.~R. {Liu}, ``Rate-energy region of {SWIPT} for
  {MIMO} broadcasting under nonlinear energy harvesting model,'' \emph{IEEE
  Trans. Wireless Commun.}, vol.~16, no.~8, pp. 5147--5161, Aug. 2017.

\bibitem{8315145}
P.~N. {Alevizos} and A.~{Bletsas}, ``Sensitive and nonlinear far-field {RF}
  energy harvesting in wireless communications,'' \emph{IEEE Trans. Wireless
  Commun.}, vol.~17, no.~6, pp. 3670--3685, Jun. 2018.

\bibitem{6613706}
A.~N. {Parks}, A.~P. {Sample}, Y.~{Zhao}, and J.~R. {Smith}, ``A wireless
  sensing platform utilizing ambient {RF} energy,'' in \emph{Proc. IEEE Radio
  Wireless Symp. (RWS)}, Austin, Tx, USA, Jan. 2013, pp. 154--156.

\bibitem{9042882}
Y.~{Hu}, M.~{Chen}, and W.~{Saad}, ``Joint access and backhaul resource
  management in satellite-drone networks: A competitive market approach,''
  \emph{IEEE Trans. Wireless Commun.}, vol.~19, no.~6, pp. 3908--3923, 2020.

\bibitem{8941314}
Z.~{Yang}, W.~{Xu}, and M.~{Shikh-Bahaei}, ``Energy efficient {UAV}
  communication with energy harvesting,'' \emph{IEEE Trans. Veh. Technol.},
  vol.~69, no.~2, pp. 1913--1927, Feb. 2020.

\bibitem{YOU20091879}
\BIBentryALTinterwordspacing
F.~You, P.~M. Castro, and I.~E. Grossmann, ``Dinkelbach's algorithm as an
  efficient method to solve a class of {MINLP} models for large-scale cyclic
  scheduling problems,'' \emph{Comput. Chem. Eng.}, vol.~33, no.~11, pp.
  1879--1889, 2009. [Online]. Available:
  \url{http://www.sciencedirect.com/science/article/pii/S0098135409001367}
\BIBentrySTDinterwordspacing

\bibitem{matsui92ananalysis}
T.~Matsui, Y.~Saruwatari, and M.~Shigeno, ``An analysis of dinkelbach’s
  algorithm for 0--1 fractional programming problems dept,'' \emph{Math. Eng.
  Inf. Phys., Univ. Tokyo, Japan, METR92-14}, 1992.

\bibitem{doi:10.1287/mnsc.22.8.868}
\BIBentryALTinterwordspacing
S.~Schaible, ``Fractional programming. {II}, on dinkelbach's algorithm,''
  \emph{Manage. Sci.}, vol.~22, no.~8, pp. 868--873, 1976. [Online]. Available:
  \url{https://doi.org/10.1287/mnsc.22.8.868}
\BIBentrySTDinterwordspacing

\bibitem{stancu2012fractional}
I.~M. Stancu-Minasian, \emph{Fractional programming: theory, methods and
  applications}.\hskip 1em plus 0.5em minus 0.4em\relax Springer Science \&
  Business Media, 2012, vol. 409.

\bibitem{doi:10.1287/mnsc.13.7.492}
\BIBentryALTinterwordspacing
W.~Dinkelbach, ``On nonlinear fractional programming,'' \emph{Manage. Sci.},
  vol.~13, no.~7, pp. 492--498, 1967. [Online]. Available:
  \url{https://doi.org/10.1287/mnsc.13.7.492}
\BIBentrySTDinterwordspacing

\bibitem{8314727}
K.~{Shen} and W.~{Yu}, ``Fractional programming for communication
  systems—part {I}: Power control and beamforming,'' \emph{IEEE Trans. Signal
  Process.}, vol.~66, no.~10, pp. 2616--2630, May 2018.

\bibitem{boyd2004convex}
S.~Boyd and L.~Vandenberghe, \emph{Convex Optimization}.\hskip 1em plus 0.5em
  minus 0.4em\relax Cambridge University press, 2004.

\bibitem{doi:10.1002/nav.20053}
H.~W. Kuhn, ``The hungarian method for the assignment problem,'' \emph{Naval
  Research Logistics (NRL)}, vol.~52, no.~1, pp. 7--21, 2005.

\bibitem{yuille2002concave}
A.~L. Yuille and A.~Rangarajan, ``The concave-convex procedure ({CCCP}),'' in
  \emph{Advances in neural information processing systems}, 2002, pp.
  1033--1040.

\bibitem{kershner1939number}
R.~Kershner, ``The number of circles covering a set,'' \emph{American Journal
  of mathematics}, vol.~61, no.~3, pp. 665--671, 1939.

\bibitem{jain1999throughput}
R.~Jain, A.~Durresi, and G.~Babic, ``Throughput fairness index: An
  explanation,'' in \emph{ATM Forum contribution}, vol.~99, no.~45, 1999.

\bibitem{1658226}
{Wei Yu} and R.~{Lui}, ``Dual methods for nonconvex spectrum optimization of
  multicarrier systems,'' \emph{IEEE Trans. Commun.}, vol.~54, no.~7, pp.
  1310--1322, Jul. 2006.

\bibitem{8424210}
X.~{Kang}, Y.-C. {Liang}, and J.~{Yang}, ``Riding on the primary: A new
  spectrum sharing paradigm for wireless-powered {IoT} devices,'' \emph{IEEE
  Trans. Wireless Commun.}, vol.~17, no.~9, pp. 6335--6347, Sep. 2018.

\bibitem{4786456}
X.~{Kang}, Y.-C. {Liang}, A.~{Nallanathan}, H.~K. {Garg}, and R.~{Zhang},
  ``Optimal power allocation for fading channels in cognitive radio networks:
  Ergodic capacity and outage capacity,'' \emph{IEEE Trans. Wireless Commun.},
  vol.~8, no.~2, pp. 940--950, Feb. 2009.

\bibitem{6661329}
D.~W.~K. {Ng}, E.~S. {Lo}, and R.~{Schober}, ``Wireless information and power
  transfer: Energy efficiency optimization in {OFDMA} systems,'' \emph{IEEE
  Trans Wireless Commun}, vol.~12, no.~12, pp. 6352--6370, Dec. 2013.

\end{thebibliography}
\end{document}